\documentclass[a4paper,11pt]{article}
\usepackage{jcappub} 
\usepackage{lineno}

\title{\boldmath Distinguishing scale-dependent Planck stars from renormalization group improved Schwarzschild black holes by Gravitational waves}

\author{Li Huang}
\affiliation{Purple Mountain Observatory, Chinese Academy of Sciences,\\
Nanjing 210023, China}
\affiliation{School of Astronomy and Space Science, University of Science and Technology of China,\\
Hefei 230026, China}

\emailAdd{lihuang@pmo.ac.cn}

\abstract{
Extreme mass-ratio inspirals (EMRIs), consisting of a stellar-mass black hole orbiting a supermassive black hole, are among the primary targets for future space-based gravitational wave detectors. By analyzing the emitted gravitational wave signals, we can probe the nature of compact objects in the strong-field region. To achieve this, we examine the effects of gravitational radiation. In this work, we base our calculations on the general relativistic Schwarzschild background and calculate the energy and angular momentum fluxes of gravitational waves. We perform a theoretical analysis of the equations of motion and the orbital evolution equations for EMRIs. The gravitational waveforms generated by the different periodic orbits of timelike test particles around scale-dependent Planck stars or renormalization group improved Schwarzschild black holes are investigated using both time-domain and frequency-domain methods. The time-domain method employs the ``analytic kludge" (AK) approach, while the frequency-domain method utilizes the discrete Fourier transform. We calculate the characteristic strain of the corresponding gravitational waves and compare them with the sensitivity curves of both ground-based and space-based detectors. These gravitational wave sensitivity curves can be experimentally tested for both spacetimes considered. Additionally, we use two approximate methods--the large eccentricity (EL) method and the small eccentricity (ES) method--to study the orbital evolution of EMRIs and compare the results with equatorial orbits derived from geodesic equations. Our findings will contribute to a deeper understanding of the nature of spacetime.}

\begin{document}
\maketitle
\flushbottom

\section{Introduction}
\label{sec:intro}
Since its inception in 1915~\cite{Einstein1915SPAW844}, Einstein's General Theory of Relativity has undergone extensive validation in weak-field regimes for more than half a century, primarily through tests conducted within the solar system. However, since 2015, breakthroughs in gravitational wave astronomy and black hole imaging have enabled direct probes of strong-field gravity, fundamentally expanding the scope of general relativity testing. Although the gravitational redshift of white dwarfs could theoretically serve as an important test of general relativity, observational and astrophysical challenges make it more difficult to measure accurately compared to the more direct and powerful tests within the solar system~\cite{Holberg2010JHA41.41}. The discovery of the first binary pulsar by Russell Hulse and Joseph Taylor in the summer of 1974~\cite{Hulse1975ApJ195.L51} provided the physics community with an entirely new testing ground.
For the first time, aspects of the gravitational interaction between strongly self-gravitating objects, particularly two neutron stars~\cite{Clark1977ApJ215.311,Damour1981PLA87.81,Damour1985AIHPA43.107}, were studied. The Hulse-Taylor Pulsar binaries (PSRB~1913+16) provide a unique way to test the theory of gravity and are important for opening a cosmic window on gravitational waves. These systems not only aid in the construction of ground-based gravitational wave observatories but also contribute to the detection of very low-frequency gravitational waves through pulsar timing arrays~\cite{Perera2019MNRAS490.4666,Moore2015CQG32.055004}.
With modern advancements in observational technology, general relativity is now being tested in more extreme environments, such as black hole binaries~\cite{LIGO2016PRL116.061102,LIGO2016PRL116.241103,LIGO2017PRL118.221101,LIGO2017ApJ851.L35,LIGO2017PRL119.141101}, neutron stars~\cite{LIGO2017PRL119.161101,Agazie2023ApJL951.L8,LIGO2022PRD106.102008,Bodaghee2023ApJ951.37}, and through gravitational waves. These tests offer a deeper and more rigorous understanding of how gravitational waves behave under stronger gravitational fields and higher velocities, pushing the boundaries of our knowledge and potentially uncovering new physics.

Gravitational waves are among the most important theoretical predictions of general relativity~\cite{Weinstein2016arXiv1602.04040,Einstein1916AnP354.769,Einstein1916SPAW688,Einstein1918SPAW154,
Eddington1922RSPSA102.268,Eddington1923Book}. Nevertheless, detecting them remains extremely challenging. The effectiveness of detection capabilities is not only limited by the sensitivity of the hardware but also depends on the accuracy of the theoretical models used to describe the wave source. EMRIs represent an important class of gravitational wave sources~\cite{Hughes2001CQG18.4067,Tinto2005LRR8.4,Folkner1998AIPC456,Babak2017PRD95.103012,Fan2020PRD102.063016}. These systems consist of a central supermassive black hole (with a mass of approximately $10^{6}-10^{7}M_{\odot}$) and a smaller stellar-mass object (with a mass of approximately $1-10M_{\odot}$) orbiting around it. EMRIs are considered among the most promising sources of gravitational waves for low-frequency space-based detectors, such as the Laser Interferometer Space Antenna (LISA)~\cite{Amaro-Seoane2017arXiv170200786,Maselli2022NA6.464,Robson2019CQG36.105011,Amaro-Seoane2012CQG29.124016}, TianQin~\cite{Luo2016CQG33.035010,Liu2020PRD101.103027}, Big Bang Observer (BBO)~\cite{Cutler2006PRD73.042001}, Decihertz Interferometer Gravitational Wave Observatory (DECIGO)~\cite{DECIGO2021PTEP2021.05A105}, and Taiji~\cite{Ruan2020IJMPA35.2050075}. The gravitational waves emitted by EMRIs are closely dependent on the orbital dynamics of the smaller celestial object~\cite{Glampedakis2005CQG22.S605}. Studying the special orbits of timelike test particles around renormalization group improved Schwarzschild black holes, as well as those around scale-dependent Planck stars, will help identify characteristic gravitational wave signals in EMRIs. Consequently, analyzing EMRIs requires accurate modeling of the corresponding gravitational waveforms~\cite{Barack2004PRD69.082005,Babak2007PRD75.024005,Chua2015CQG32.232002,Chua2017PRD96.044005,Rahman2023PRD107.024006,
Rahman2024JCAP01.035,Speri2023arXiv2307.12585,Zi2023PRD107.023005,Zi2023PRD108.024018,Katz2021PRD104.064047,Yang2025JCAP01.091}. Accurately and efficiently calculating the waveforms of EMRIs is a key scientific challenge. This is necessary for detecting such signals from observational data and for precisely determining the parameters of the wave sources. The numerical solution of perturbation equations primarily involves two approaches: time-domain methods and frequency-domain methods.

Time-domain methods obtain numerical results by directly solving partial differential equations, which include source terms~\cite{Sundararajan2007PRD76.104005,Sundararajan2008PRD78.024022}. In contrast, frequency-domain methods use discrete Fourier transforms and separation of variables to convert the equations into ordinary differential equations. Subsequently, Green's functions are constructed to solve these equations~\cite{Hughes2000PRD61.084004,Drasco2006PRD73.024027}. The frequency-domain method is the primary approach for calculating the waveforms of EMRIs. Since EMRIs contain a wealth of wave source information, accurate data analysis can be achieved by calculating the waveform using the post-Newtonian approximation~\cite{Poisson1995PRD51.5753}, which expands the equations of motion in powers of $v/c$. However, a limitation of this method is that the post-Newtonian approximation converges poorly when $v/c>0.3$~\cite{Amaro-Seoane2007CQG24.R113}. Despite this, the AK approximation~\cite{Barack2004PRD69.082005,Babak2007PRD75.024005} is advantageous in terms of computational efficiency while still maintaining the gravitational wave signal characteristics of EMRIs. As a result, it remains popular in the study of EMRIs.

This paper explores waveform calculations for gravitational waves in the context of EMRIs. The primary approach combines both time-domain and frequency-domain methods ~\cite{Gair2007CQG24.1145,Gair2008CQG25.184031} to investigate the gravitational waveforms produced by the different periodic orbits of a small body (timelike test particle) around scale-dependent Planck stars or renormalization group improved Schwarzschild black holes. The study of zoom-whirl orbits around these scale-dependent Planck stars, emerging from renormalization group improved Schwarzschild black holes, is motivated by the fact that these systems are promising sources for low-frequency gravitational wave observatories, such as LISA~\cite{Amaro-Seoane2017arXiv170200786,Maselli2022NA6.464,Robson2019CQG36.105011,Amaro-Seoane2012CQG29.124016}. For a static spacetime background and geodesic particle trajectories, gravitational waveforms can be systematically derived to highlight the imprint of zoom-whirl orbital dynamics. Leveraging the methodologies outlined in refs.~\cite{Peters1964PR136.1224,Junker1992MNRAS254.146,Forseth2016PRD93.064058,Gopakumar1997PRD56.7708,
Gopakumar1998PhRvD57.6562,Ashtekar2002PRL89.261101,Hasselmann1963JFM15.385,Maggiore2007gwte.book,Poisson2008CQG25.209002,
Thorne1980RMP52.299,Blanchet2002PRD65.064005,Blanchet2005PRD71.129903,Boyle2008PRD78.104020}, we compute the energy and angular momentum fluxes emitted in gravitational waves. Here, the energy flux can be expressed in terms of the source's quadrupole moment and the properties of gravitational emission. Similarly, the angular momentum carried away by the gravitational waves lead to a change in the system's orbital angular momentum, which ultimately result in the merger of black hole or its transition to a different orbital configuration. It is common practice to describe both the sensitivity of a gravitational wave detector and the strength of a gravitational wave source using a sensitivity plot~\cite{Thrane2013PRD88.l24032,Torres2011heep,Moore2015CQG32.015014}. The amplitude of a gravitational wave is represented by strain, a dimensionless quantity. The characteristic strain provides a way to express the strength of gravitational waves across different frequencies, typically used to compare with detector sensitivities. To ensure the accuracy of future detections, it is essential to quantify both the sensitivity of the instrument and the strength of its target signal. Three commonly used parametrizations based on strain are typically employed when discussing the strength of gravitational wave sources and detector sensitivity: characteristic strain, power spectral density, and the spectral energy density. In this paper, we focus on the characteristic strain of timelike test particles around scale-dependent Planck stars or renormalization group improved Schwarzschild black holes. Simultaneously, we compare this characteristic strain with the sensitivity curves of various detectors, including LISA~\cite{Robson2019CQG36.105011}, eLISA~\cite{Amaro-Seoane2012CQG29.124016}, TianQin~\cite{Luo2016CQG33.035010}, BBO~\cite{Cutler2006PRD73.042001}, DECIGO~\cite{DECIGO2021PTEP2021.05A105}, European Pulsar Timing Array (EPTA)~\cite{Antoniadis2023A&A678.A50}, International Pulsar Timing  Array (IPTA)~\cite{Hobbs2010CQG27.084013}, the Square Kilometre Array (SKA)~\cite{Braun2015SKA174}, Laser Interferometer Gravitational-wave Observatory (LIGO)~\cite{LIGO2017CQG34.044001}, Advanced LIGO (aLIGO)~\cite{LIGO2021PRX11.021053}, and LIGO~$\mathrm{A^{+}}$~\cite{Tso2022Aplus}. Besides, we present a precise solution for the orbital evolution of EMRIs, based on the EL and ES methods, and compare the results with equatorial orbits derived from geodesic equations.

The outline of this paper is as follows: In section~\ref{sec:2}, we provide a brief review the metric and equations of motion, followed by the calculation of orbital frequencies. In section~\ref{sec:3}, we discuss the orbital evolution equations for EMRIs, focusing on Keplerian parameters and their applications. This section also includes discussions on the EL and ES methods. In section~\ref{sec:4}, we compute the energy and angular momentum fluxes of gravitational waves. Our primary focus is on analytic solutions and their applications, specifically investigating gravitational waveforms generated by the different periodic orbits of timelike test particles around scale-dependent Planck stars or renormalization group improved Schwarzschild black holes. Both time-domain and frequency-domain methods are employed for this analysis. We also provide the corresponding characteristic strain and compared them with those from various gravitational wave detectors, including  LISA, eLISA, TianQin, BBO, DECIGO, EPTA, IPTA, SKA, LIGO, aLIGO, LIGO~$\mathrm{A^{+}}$. Finally, we conclude the paper in section~\ref{sec:5}. Throughout this paper, we adopt the metric convention $(-,+,+,+)$ and use the units $G=c=\hbar=1$.

\section{Equations of motion for extreme mass-ratio inspirals}
\label{sec:2}
\subsection{Metric}
\label{Sec:2.1}
The metric for the scale-dependent Planck stars and the renormalization group improved Schwarzschild black holes in the Schwarzschild gauge can be expressed as in refs.~\cite{Bonanno2000PRD62.043008,Scardigli2023PRD107.104001,Bonanno2002PRD65.043508,Huang2024PRD109.l24005}, as given in
\begin{eqnarray}
\label{metric3}
\mathrm{d}s^2&=&-f(r)\mathrm{d}t^2+\frac{1}{f(r)}\mathrm{d}r^2+r^2\mathrm{d}\Omega^2.
\end{eqnarray}
For a static, spherical system, we adopt the convention of the metric with signature $(-,+,+,+)$. In metric \eqref{metric3},  $\mathrm{d}\Omega^2=\mathrm{d}\theta^2+\sin^2\theta\mathrm{d}\phi^2$, which represents the metric on the two-sphere. The metric component in eq.~\eqref{metric3} is given by
\begin{eqnarray}
\label{daomo}
f(r)&=&1-\frac{2Gm_{\bullet}}{c^2r}\left(1+s|\tilde{\omega}|\frac{G\hbar}{c^3r^2}+\gamma s|\tilde{\omega}| \frac{G^2\hbar m_{\bullet}}{c^5r^3}\right)^{-1}.
\nonumber \\
\end{eqnarray}
Introducing a positive dimensionless parameter $\Omega\equiv|\tilde{\omega}|(\hbar c)/(Gm^{2}_{\bullet})\geq0$, eq.~\eqref{daomo} can be rewritten as
\begin{eqnarray}
\label{metricfh}
f(r)&=&1-\frac{2m_{\bullet}}{r}\left(1+s\Omega \frac{m_{\bullet}^2}{r^2}+\gamma s\Omega \frac{m_{\bullet}^3}{r^3}\right)^{-1},
\end{eqnarray}
by adopting units in which $G=c=\hbar=1$, unless stated otherwise. In the metric component \eqref{metricfh}, $\gamma$ is a positive dimensionless parameter that corresponds to the cutoff of the associated distance scale~\cite{Bonanno2000PRD62.043008,Lambiase2022PhRvD105.124054}. Additionally, $\Omega$ arises from nonperturbative renormalization group theory and represents the dimensionless coupling parameter~\cite{Bonanno2000PRD62.043008,Scardigli2023PRD107.104001,Lambiase2022PhRvD105.124054}. The sign parameter $s$ is used to differentiate between scale-dependent Planck stars and renormalization group improved Schwarzschild black holes. For instance, $s=-1$ corresponds to scale-dependent Planck stars, while $s=1$ represents renormalization group improved Schwarzschild black holes. For convenience, we define
$x\equiv r/m_{\bullet}$. The metric component \eqref{metricfh} then becomes
\begin{eqnarray}
\label{FFF}
f(x)&=&1-\frac{2}{x}(1+s\Omega x^{-2}+\gamma s\Omega x^{-3})^{-1}.
\end{eqnarray}
Where, for scale-dependent Planck stars,
\begin{eqnarray}
\label{Omegafu}
\Omega&=&\lambda_{-}\Omega_{-},
\nonumber \\
\Omega_{-}&=&-\frac{27}{8}{\gamma}^2-\frac{9}{2}\gamma+\frac{1}{2}-\frac{1}{8}\sqrt{(\gamma+2)(9\gamma+2)^3}.
\end{eqnarray}
For renormalization group improved Schwarzschild black holes,
\begin{eqnarray}
\label{Omegazh}
\Omega&=&\lambda_{+}\Omega_{+},
\nonumber \\
\Omega_{+}&=&-\frac{27}{8}{\gamma}^2-\frac{9}{2}\gamma+\frac{1}{2}+\frac{1}{8}\sqrt{(\gamma+2)(9\gamma+2)^3}.
\end{eqnarray}
Note that $f(x)$ is a function of $\gamma$ and $\Omega$, and when both are zero, the metric \eqref{metric3} will be identical to Schwarzschild black holes.

The (in)existence of the event horizon(s) in Planck stars and black holes has been examined in previous works~\cite{Bonanno2000PRD62.043008,Bonanno2002PRD65.043508,Scardigli2023PRD107.104001,Huang2024PRD109.l24005} through the condition $f(x)=0$. It has been found that one event horizon always exists for scale-dependent Planck stars $(\lambda_{-}<0)$, while for renormalization group improved Schwarzschild black holes, there are three possible scenarios: one $(\lambda_{+}=1)$, two $(0<\lambda_{+}<1)$, or even no event horizon(s) $(\lambda_{+}>1)$, depending on the values of $\gamma$ and $\Omega$. A detailed discussion of the (in)existence of horizons for these two spacetimes can be found in refs.~\cite{Bonanno2000PRD62.043008,Bonanno2002PRD65.043508,Scardigli2023PRD107.104001,Huang2024PRD109.l24005}.

\subsection{Equations of motion}
\label{Sec:2.2}
For a timelike test particle in a gravitational field, its Lagrangian in the equatorial plane (where $\theta=\pi/2$) can be expressed as
\begin{eqnarray}
\label{lagelrt}
2\mathcal{L}&=&-f(x)\dot{t}^2+\frac{1}{f(x)}\dot{x}^2+x^2\dot{\phi}^2=-1,
\end{eqnarray}
where, ``$\cdot$" denotes the derivative with respect to an affine parameter. Two conserved quantities can be defined: the energy and angular momentum per unit mass of the particle, which can be written as
\begin{eqnarray}
\label{lageEr}
E&=&f(x)\dot{t},
\\
\label{lagelr}
l&=&x^2\dot{\phi}.
\end{eqnarray}
Furthermore, by separating the variables, the Hamilton-Jacobi equation yields an equation for $x$, which can be expressed as follows:
\begin{eqnarray}
\label{drdtau}
\dot{x}^2&=&E^2-f(x)\Big(1+\frac{l^2}{x^2}\Big).
\end{eqnarray}
The radial distance $x$ can be written in the following form:
\begin{eqnarray}
\label{rchi}
x&=&\frac{p}{1+e\cos{\chi}},
\end{eqnarray}
where, $p$ is the dimensionless semilatus rectum, $e$ is the orbital eccentricity, and $\chi$ is the relativity true anomaly. At $\chi=0$ and $\chi=\pi$, the distances of the orbit from the perihelion and aphelion are
\begin{eqnarray}
\label{jinxin}
x_\mathrm{pe}&=&\frac{p}{1+e},
\\
\label{yuanxin}
x_\mathrm{ap}&=&\frac{p}{1-e}.
\end{eqnarray}
By combining eq.~\eqref{rchi} and differentiating it, the relativistic true anomaly $\chi$ varies with the affine parameter as
\begin{eqnarray}
\label{chitau}
\dot{\chi}&=&\frac{(1+e\cos{\chi})^2}{ep|\sin{\chi}|}\dot{x}.
\end{eqnarray}
It should be noted that the radial velocity of small objects in EMRIs is zero at the periastron and apoastron, i.e., $\dot{x}=0$ at $x=x_\mathrm{pe}$ and $x_\mathrm{ap}$. For bound orbits, the eccentricity lies in the range $0<e<1$. Therefore, $\dot{\chi}$ maintains the same sign as $\dot{x}$, implying that $\dot{\chi}\geq 0$. This means that $\chi$ increases monotonically with the affine parameter. Thus, the equations of motion for scale-dependent Planck stars and renormalization group improved Schwarzschild black holes can be reduced to
\begin{eqnarray}
\label{tdchi}
\dot{\chi}&=&\frac{(1+e\cos{\chi})^2}{ep|\sin{\chi}|}\sqrt{E^2-f(x)\Big(1+\frac{l^2}{x^2}\Big)}.
\end{eqnarray}
Since $\dot{x}=0$ at $x_\mathrm{pe}$ and $x_\mathrm{ap}$,
we can derive from eq.~\eqref{drdtau} the relationship between the dimensionless semilatus rectum $p$, the orbital eccentricity $e$, and the energy
\begin{eqnarray}
\label{eeEE}
E&=&\frac{\sqrt{f(x_\mathrm{ap})f(x_\mathrm{pe})}\sqrt{x_\mathrm{ap}^2-x_\mathrm{pe}^2}}{\sqrt{f(x_\mathrm{pe})x_\mathrm{ap}^2-f(x_\mathrm{ap})x_\mathrm{pe}^2}},
\end{eqnarray}
as well as the angular momentum
\begin{eqnarray}
\label{anANL}
l&=&\frac{\sqrt{x_\mathrm{ap}^2x_\mathrm{pe}^2}\sqrt{f(x_\mathrm{ap})-f(x_\mathrm{pe})}}{\sqrt{f(x_\mathrm{pe})x_\mathrm{ap}^2-f(x_\mathrm{ap})x_\mathrm{pe}^2}}.
\end{eqnarray}
Expanding the energy $E$ and angular momentum $l$ in terms of dimensionless semilatus rectum $p$, while retaining the $p^{-4}$ term, only the lower-order terms are considered, and the effects of higher-order terms are neglected. The specific expression can then be written as a two-part contribution:
\begin{eqnarray}
\label{fEzong}
E&=&E_{\mathrm{Schw}}+E_{\mathrm{s\Omega}}.
\end{eqnarray}
Where $E_{\mathrm{Schw}}$ represents the contribution from Schwarzschild black holes, given by
\begin{eqnarray}
\label{fEsch}
E_{\mathrm{Schw}}&=&1-\frac{1}{2p}(1-e^2)+\frac{3}{8p^2}(1-e^2)^2+\frac{1}{16p^3}(1-e^2)^2(27+5e^2)
\nonumber \\
&&+\frac{1}{128p^4}(1-e^2)^2(27+5e^2)(25+7e^2)+\mathcal{O}(p^{-5}),
\end{eqnarray}
and $E_{\mathrm{s\Omega}}$ represents the leading term from scale-dependent Planck stars or renormalization group improved Schwarzschild black holes, given by
\begin{eqnarray}
\label{fEsOmega}
E_{\mathrm{s\Omega}}&=&-\frac{1}{2p^3}s\Omega (1-e^2)^2-\frac{1}{4p^4}s\Omega(1-e^2)^2(13+3e^2+4\gamma)+\mathcal{O}(p^{-5}).
\end{eqnarray}
When $\gamma=\Omega=0$, eq.~\eqref{fEzong} reduces to the classical Schwarzschild case. It should be noted that when $e=0$, we have $\dot{x}=\ddot{x}=0$, and $x_\mathrm{pe}=x_\mathrm{ap}=p$. In this case, eq.~\eqref{fEzong} corresponds to a circular orbit. Similarly, the angular momentum can be decomposed into two components:
\begin{eqnarray}
\label{fLzong}
l&=&l_{\mathrm{Schw}}+l_{\mathrm{s\Omega}}.
\end{eqnarray}
Here, $l_{\mathrm{Schw}}$ stands for the Schwarzschild black hole contribution, expressed as
\begin{eqnarray}
\label{flsch}
l_{\mathrm{Schw}}&=&\frac{1}{p^{-1/2}}+\frac{1}{2p^{1/2}}(3+e^2)+\frac{3}{8p^{3/2}}(3+e^2)^2+\frac{5}{16p^{5/2}}(3+e^2)^3+\frac{35}{128p^{7/2}}(3+e^2)^4
\nonumber \\
&&+\frac{63}{256p^{7/2}}(3+e^2)^5+\mathcal{O}(p^{-9/2}),
\end{eqnarray}
and $l_{\mathrm{s}\Omega}$ denotes the leading term for scale-dependent Planck stars or renormalization group improved Schwarzschild black holes, expressed as
\begin{eqnarray}
\label{flsOmega}
l_{\mathrm{s\Omega}}&=&-\frac{1}{2p^{3/2}}s\Omega(3+e^2)-\frac{1}{4p^{5/2}}s\Omega\Big[19+3e^4+8\gamma+e^2(26+8\gamma)\Big]
\nonumber \\
&&+\frac{1}{16p^{7/2}}s\Omega\Big[(-3-e^2)\big(87+15e^4+32\gamma+2e^2(69+32\gamma)+2(11+e^2)(1+3e^2)s\Omega\big)\Big]
\nonumber \\
&&+\mathcal{O}(p^{-9/2}).
\end{eqnarray}
The above equation is preserved up to $p^{-7/2}$, with the leading term contribution considered. when $\gamma=\Omega=0$, eq.~\eqref{fLzong} reduces to the classical Schwarzschild case.

\subsection{Orbital frequencies}
\label{Sec:2.3}
In the equatorial plane, the frequency corresponding to the polar angle ($\theta$) is negligible. For bound orbits, only the two frequencies $\Upsilon_x$ and $\Upsilon_{\phi}$, associated with $x(\chi)$ and $\chi$, remain. The time elapsed for a celestial body to move from one periastron (or apoastron) to the next periastron (or apoastron)~\cite{Fujita2009CQG26.135002} is given by
\begin{eqnarray}
\label{omtzong}
\Lambda_t&&=\int_0^{t_0}\mathrm{d}t=\int_0^{2\pi}\frac{\mathrm{d}t}{\mathrm{d}\chi}\mathrm{d}\chi
\nonumber \\
&&=\Lambda_{t(\mathrm{Schw})}+\Lambda_{t(\mathrm{s\Omega})},
\end{eqnarray}
where
\begin{eqnarray}
\label{omtone}
\Lambda_{t(\mathrm{Schw})}&=&\frac{2\pi}{(\sqrt{1-e^2})^3}p^{3/2}+\frac{3\pi}{\sqrt{1-e^2}}p^{1/2}-\frac{3\pi}{4\sqrt{1-e^2}}(-17+e^2-4\sqrt{1-e^2})p^{-1/2}
\nonumber \\
&&-\frac{\pi}{8\sqrt{1-e^2}}\big[-431+e^4-244\sqrt{1-e^2}
+2e^2(23+6\sqrt{1-e^2})\big] p^{-3/2}
\nonumber \\
&&-\frac{3\pi}{64\sqrt{1-e^2}}
\big[e^6+e^4(29+8\sqrt{1-e^2})-27(179+184\sqrt{1-e^2})
\nonumber \\
&&+e^2(707+328\sqrt{1-e^2})\big] p^{-5/2}+\mathcal{O}(p^{-7/2}),
\nonumber \\
\end{eqnarray}
and
\begin{eqnarray}
\label{omttwo}
\Lambda_{t(\mathrm{s\Omega})}&=&-\frac{3\pi s\Omega}{\sqrt{1-e^2}}p^{-1/2}-\frac{\pi s\Omega}{2\sqrt{1-e^2}}
\big[24\sqrt{1-e^2}+51-3e^2++4\gamma(3+\sqrt{1-e^2})\big]p^{-3/2}
\nonumber \\
&&+\frac{3\pi s\Omega}{8\sqrt{1-e^2}}\big[e^4-7(57+52\sqrt{1-e^2})-104\gamma+34s\Omega+2e^2(7+6\sqrt{1-e^2}
\nonumber \\
&&-12\gamma-s\Omega)+8\sqrt{1-e^2}(-11\gamma+3s\Omega)\big]p^{-5/2}+\mathcal{O}(p^{-7/2}).
\end{eqnarray}
The corresponding variation of the azimuthal angle $\phi$ is
\begin{eqnarray}
\label{ophizong}
\Lambda_{\phi}&&=\int_0^{\phi_0}\mathrm{d}\phi=\int_0^{2\pi}\frac{\mathrm{d}\phi}{\mathrm{d}\chi}\mathrm{d}\chi
\nonumber \\
&&=\Lambda_{\phi(\mathrm{Schw})}+\Lambda_{\phi(\mathrm{s\Omega})},
\end{eqnarray}
where
\begin{eqnarray}
\label{ophione}
\Lambda_{\phi(\mathrm{Schw})}&=&2\pi+\frac{6\pi}{p}+\frac{3\pi}{2p^2}(18+e^2)+\frac{45\pi}{2p^3}(6+e^2)+\frac{105\pi}{32p^4}(216+72e^2+e^4)
\nonumber \\
&&+\mathcal{O}(p^{-5}),
\end{eqnarray}
and
\begin{eqnarray}
\label{ophitwo}
\Lambda_{\phi(\mathrm{s\Omega})}&=&-\frac{6\pi s\Omega}{p^2}-\frac{\pi s\Omega}{p^3}\big[56+12\gamma+3e^2(4+\gamma)\big]-\frac{\pi s\Omega}{4p^4}\big[3e^4(3+\gamma)
+4(423+102\gamma
\nonumber \\
&&-29s\Omega)+6e^2(113+42\gamma-7s\Omega)\big]+\mathcal{O}(p^{-5}).
\end{eqnarray}
Similarly, the system reduces to the classical Schwarzschild case when $\gamma=\Omega=0$.

Angular velocity plays a key role in describing the dynamics of celestial orbits~\cite{Newton1687book,Brouwer1961Book,Laskar1989Nat338.237,Mayor1995Nat378.355,Liu2024JCAP10.056}. There exists a gauge-invariant quantity for angular velocity, which is a conserved quantity. Its Lagrangian and equations of motion are invariant under canonical transformations, making them essential for the construction of gauge theory.
Furthermore, since angular velocity is gauge-invariant, it implies that the angular velocity of a system remains constant in the absence of external torques. This property allows the position of a celestial body to be expressed in terms of the frequency of its circular orbit, such as the familiar motion of a planet around a star. However, it is important to note that while frequency can describe the motion of a celestial body, it is not equivalent to position. Frequency is merely a parameter that indicates the rate at which an object moves, whereas position refers to the specific coordinates of a celestial body in space. Angular velocity is commonly used in astronomy to describe the motion of planets, satellites, and other celestial bodies. In certain special cases, such as when the gravitational perturbations from other celestial bodies are neglected, the angular velocities of planets can be considered approximately conserved. Likewise, angular velocity is used in orbital dynamics to calculate orbital periods, orbital radii, and other related parameters. In practical applications, such as satellite navigation systems, the frequency of a satellite's motion around Earth can be used to accurately determine its position. Therefore, angular velocity is a crucial parameter in describing the motion of celestial bodies.

Continuing from the previous expressions and context, the average angular velocity of $x$ and $\phi$ in each cycle can be expressed as
\begin{eqnarray}
\label{jiaot}
\Upsilon_{x}&=&\frac{2\pi}{\Lambda_{t}},
\\
\label{jiaophi}
\Upsilon_{\phi}&=&\frac{\Lambda_{\phi}}{\Lambda_t}.
\end{eqnarray}
The following expansion can be obtained by calculation
\begin{eqnarray}
\label{jomer}
\Upsilon_{x}&=&\frac{(1-e^2)^{3/2}}{p^{3/2}}-\frac{3(1-e^2)^{5/2}}{2p^{5/2}}-\frac{3(1-e^2)^{5/2}}{8p^{7/2}}(11+5e^2+4\sqrt{1-e^2}-4s\Omega)
\nonumber \\
&&+\mathcal{O}(p^{-9/2}),
\\
\label{jomephi}
\Upsilon_{\phi}&=&\frac{(1-e^2)^{3/2}}{p^{3/2}}
+\frac{3(1-e^2)^{1/2}}{2p^{5/2}}(1-e^4)+\frac{3(1-e^2)^{3/2}}{8p^{7/2}}(5e^4+20e^2-4\sqrt{1-e^2}(1-e^2)
\nonumber \\
&&-4e^2 s\Omega-4s\Omega+13)+\mathcal{O}(p^{-9/2}).
\end{eqnarray}
Therefore, the frequency of the periastron precession can be determined as
\begin{eqnarray}
\label{jdpl}
\frac{\mathrm{d}\hat{\Upsilon}}{\mathrm{d}t}&&=\Upsilon_{\phi}-\Upsilon_{x}
\nonumber \\
&&=\frac{3(1-e^2)^{3/2}}{p^{5/2}}+\frac{3(1-e^2)^{3/2}}{4p^{7/2}}(12+7e^2-4s\Omega)+\mathcal{O}(p^{-9/2}).
\end{eqnarray}
Here, the parameter $\gamma$ appears in eqs.~\eqref{jomer}--\eqref{jdpl} in the term $\mathcal{O}(p^{-9/2})$, which will not be expanded further, as the expression is too lengthy. It is also straightforward to verify that when $\gamma=\Omega=0$, the precession frequency energy reduces to the classical Schwarzschild case.

\section{Orbital evolution equations for extreme-mass ratio inspirals}
\label{sec:3}
We can use the Keplerian frequency to describe orbital motion. From the eqs.~\eqref{jomer} and \eqref{jomephi}, it is evident that the expressions involve two orbital frequencies, $\Upsilon_{x}$ and $\Upsilon_{\phi}$.
According to the traditional numerical integration method, we assume $\Upsilon_x=2\pi \nu_x$ and $\Upsilon_{\phi}=2\pi \nu_{\phi}$. The renormalization group-improved Schwarzschild black holes and scale-dependent Planck stars discussed in this paper are described by a static and spherically symmetric metric. Therefore, the problem of expressing the evolution equations for geodesics in terms of Keplerian frequencies can be addressed by considering the higher-order terms of $\mathrm{d}\hat{\Upsilon}/{\mathrm{d}t}$. By considering these higher-order terms, we obtain a more precise or generalized expression to describe the behavior of geodesics. It should be noted that, since the geodesic equations in general relativity are typically nonlinear and involve complex spacetime structures, making it challenging to find exact analytical solutions. As a result, approximate methods, numerical simulations, or experimental observations are often necessary to investigate and understand the evolutionary behavior of geodesics in practical applications. In this section, we will apply two main approximation methods from ref.~\cite{Liu2024JCAP10.056} to investigate orbital motion: the EL method and the ES method. These methods will be compared with equatorial orbits derived from the geodesic equations.

\subsection{The large eccentricity method}
\label{sec:3.1}
The main idea of the EL method is to first set $\Upsilon_x=2\pi \nu_x$, and then obtain the relationship between $p$ and the radial frequency $2\pi \nu_x$ by solving the inverse function of eq.~\eqref{jomer}. The higher-order expression for $\Upsilon_x$ can be derived by substituting the radial frequency expression for $p$ into eq.~\eqref{jomer}.
Finally, by substituting the specific value of $p$ into eq.~\eqref{jdpl}, we can calculate the frequency of the near-star point progression, as well as the evolution equation of the geodesic. Specifically, the expression for $p$ is as follows:
\begin{eqnarray}
\label{rxpzhi}
p&=&\frac{1-e^2}{(2\pi\nu_x)^{2/3}}-1+e^2-\frac{1}{4}(6+10e^2+4\sqrt{1-e^2}-4s\Omega)(2\pi\nu_x)^{2/3}+\frac{5}{8}(5e^2+11
\nonumber \\
&&+4\sqrt{1-e^2}-4s\Omega)(2\pi\nu_x)^{4/3}+\mathcal{O}[(2\pi\nu_x)^2].
\end{eqnarray}
By substituting eq.~\eqref{rxpzhi} into eq.~\eqref{jdpl}, the corresponding geodesic evolution equations are given by
\begin{eqnarray}
\label{elone}
\frac{\mathrm{d}\Psi}{\mathrm{d}t}&=&\Upsilon_x=2\pi\nu_x,
\\
\label{eltwo}
\frac{\mathrm{d}\Phi}{\mathrm{d}t}&=&\Upsilon_{\phi}=\Upsilon_x+\frac{\mathrm{d}\hat{\Upsilon}}{\mathrm{d}t},
\\
\label{elthree}
\frac{\mathrm{d}\hat{\Upsilon}}{\mathrm{d}t}&=&\frac{3(2\pi\nu_x)^{5/3}}{1-e^2}+\frac{(2\pi\nu_x)^{7/3}}{4(1-e^2)^2}(66-9e^2-12s\Omega)
+\frac{(2\pi\nu_x)^3}{8(1-e^2)^2}(342+297e^2
\nonumber \\
&&+60\sqrt{1-e^2}-144s\Omega)+\mathcal{O}[(2\pi\nu_x)^{11/3}],
\end{eqnarray}
where $\Psi$ represents the mean anomaly for $\chi$, and $\Phi$ represents the mean longitude for $\phi$.

\subsection{The small eccentricity method}
Adopting similar ideas as described above, and assuming $\Upsilon_{\phi}=2\pi \nu_{\phi}$ under the ES method, the expression for $p$ can be derived by solving the inverse function of eq.~\eqref{jomephi}, as follows:
\label{sec:3.2}
\begin{eqnarray}
\label{phixpz}
p&=&\frac{1-e^2}{(2\pi\nu_{\phi})^{2/3}}+1+e^2+
\frac{1}{4(1-e^2)}\big(10e^4+30e^2-4s\Omega-4(1-e^2)\sqrt{1-e^2}-4e^2s\Omega
\nonumber \\
&&+18\big)(2\pi\nu_{\phi})^{2/3}+\frac{5(1+e^2)}{8(1-e^2)^2}(5e^4+20e^2-4(1-e^2)\sqrt{1-e^2})-4s\Omega-4e^2s\Omega
\nonumber \\
&&+13)(2\pi\nu_{\phi})^{4/3}+\mathcal{O}[(2\pi\nu_{\phi})^2].
\end{eqnarray}
By substituting eq.~\eqref{phixpz} into eq.~\eqref{jdpl}, the corresponding geodesic evolution equations for this case are given by
\begin{eqnarray}
\label{esone}
\frac{\mathrm{d}\Psi}{\mathrm{d}t}&=&\Upsilon_x=\Upsilon_{\phi}-\frac{\mathrm{d}\hat{\Upsilon}}{\mathrm{d}t},
\\
\label{estwo}
\frac{\mathrm{d}\Phi}{\mathrm{d}t}&=&\Upsilon_{\phi}=2\pi\nu_{\phi},
\\
\label{esthree}
\frac{\mathrm{d}\hat{\Upsilon}}{\mathrm{d}t}&=&\frac{3(2\pi\nu_{\phi})^{5/3}}{1-e^2}+\frac{(2\pi\nu_{\phi})^{7/3}}{4(1-e^2)^2}(6-9e^2-12s\Omega)
-\frac{(2\pi\nu_x)^3}{8(1-e^2)^3}(297e^4+849e^2
\nonumber  \\
&&-144e^2s\Omega-144s\Omega-60(1-e^2)\sqrt{1-e^2}+522)+\mathcal{O}(2\pi\nu_{\phi})^{11/3}.
\end{eqnarray}
Through numerical integration and analytical comparison of the equations, it is evident that at $\Omega=0$, eqs.~\eqref{elthree} and \eqref{esthree} can also be expanded to higher orders. This leads to the approximation of $\mathrm{d}\hat{\Upsilon}/\mathrm{d}t$ being computed to $n$ orders as: $\mathcal{O}\big[(2\pi\nu_x)^{(2n+3)/3)}\big]$ or $\mathcal{O}\big[(2\pi\nu_{\phi})^{(2n+ 3)/3)}\big]$.

\subsection{Orbital motion}
This section focuses on the orbital motion of timelike test particles around scale-dependent Planck stars and those around renormalization group improved Schwarzschild black holes, employing the EL and ES methods. The orbital behaviors are analyzed using both methods. It is important to note that when calculating the numerical solution, higher-order values of $\mathrm{d}\hat{\Upsilon}/\mathrm{d}t$ in eqs.~\eqref{elthree} or \eqref{esthree} are typically used to minimize the accumulation of errors over time. The system of  eqs.~\eqref{elone}--\eqref{elthree}, and \eqref{esone}--\eqref{esthree} is solved numerically. For comparison, we also examine the corresponding orbits using the equations of motion in section~\ref{Sec:2.2}. Since the analysis is restricted to the equatorial plane, we set $\theta=\pi/2$. High-precision numerical results can be obtained using eqs.~\eqref{lageEr}--\eqref{lagelr} and \eqref{tdchi}.

Using classical methods~\cite{Maggiore2007gwte.book,Poisson2008CQG25.209002,Thorne1980RMP52.299}, we derive the relationship between $\Psi$ (mean anomaly) and $u$ (eccentric anomaly), expressed as
\begin{eqnarray}
\label{chiuone}
u&=&\Psi+e\sin{u}.
\end{eqnarray}
At the initial time $t=0$, we have $\chi=0$, where $\chi$ is the relativistic true anomaly. Trigonometric identities are then employed to derive
\begin{eqnarray}
\label{chiutwo}
\chi(u)=2\arctan\Big[\Big(\frac{1+e}{1-e}\Big)^{1/2}\tan\Big(\frac{u}{2}\Big)\Big].
\end{eqnarray}
The second-order approximation of $u$ is expressed as
\begin{eqnarray}
\label{chiuthree}
u&\approx&\Psi+e\sin{\Psi}+\frac{1}{2}e^2\sin{2\Psi},
\end{eqnarray}
while $\phi$ can be written as
\begin{eqnarray}
\label{chiufour}
\phi &\approx& \chi\Big(1+\frac{\hat{\Upsilon}}{\Psi}\Big).
\end{eqnarray}
The orbits of the timelike test particles are described by the coordinates $x^i=(x\cos{\phi},x\sin{\phi},0)$, where $x$ is defined in eq.~\eqref{rchi}, $\hat{\Upsilon}$ is the mean precession angle, and the period error in the orbit can be related to $\hat{\Upsilon}$ and $\Psi$. Next, we will use both the EL and ES methods to research the orbital evolution of EMRIs.

\begin{figure*}
\centering
\includegraphics[width=1.08\textwidth]{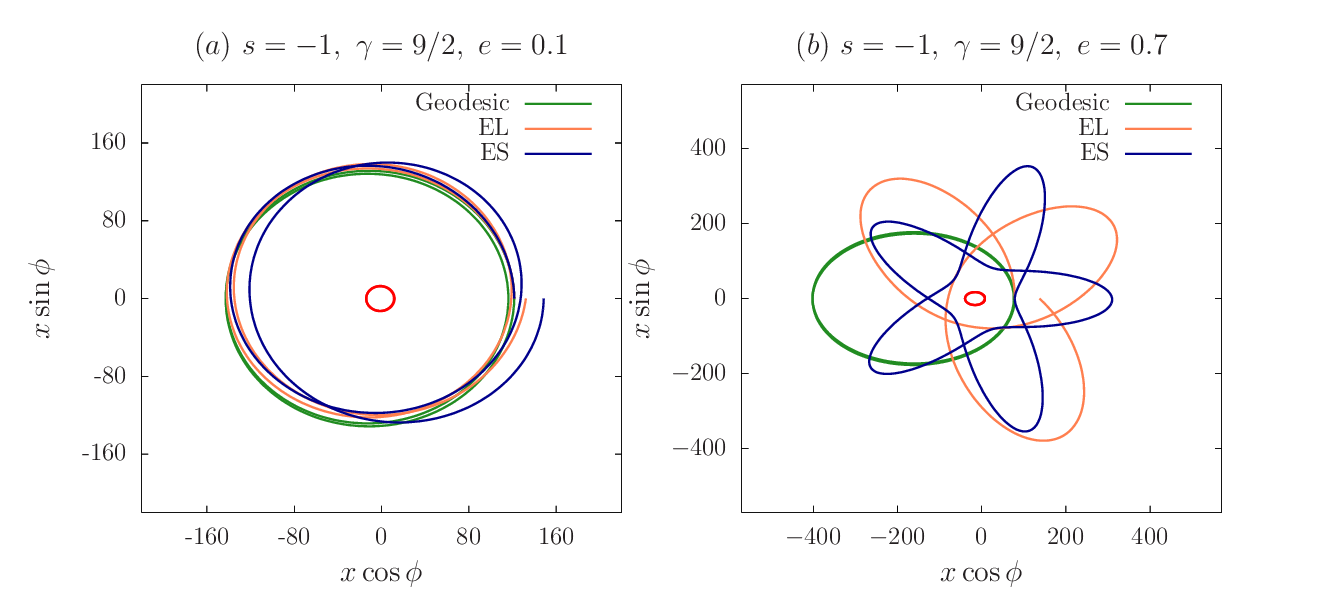}
\caption{Comparison of orbital motion for timelike test particles around scale-dependent Planck stars. Two approximate orbits from the EL (coral curve) and ES (dark-blue curve) methods are compared to the equatorial orbits (forest-green curve), which are derived from high-precision numerical solutions for eqs.~\eqref{lageEr}--\eqref{lagelr} and \eqref{tdchi}. Parameters: $M=10^7M_{\odot}$, $\gamma=9/2$,  $\lambda_{-}=-1.0$, $\Omega=176.392$, $l=8.6$, $E=0.99446821$. The frequency relation $\Upsilon_x/(2\pi)=0.2$ mHz corresponds to $\Upsilon_{\phi}/(2\pi)=0.345746$ mHz. The red curve represents the Schwarzschild case ($\gamma=\lambda_{-}=\Omega=0$).}
\label{fig:1}
\end{figure*}

\begin{figure*}
\centering
\includegraphics[width=1.08\textwidth]{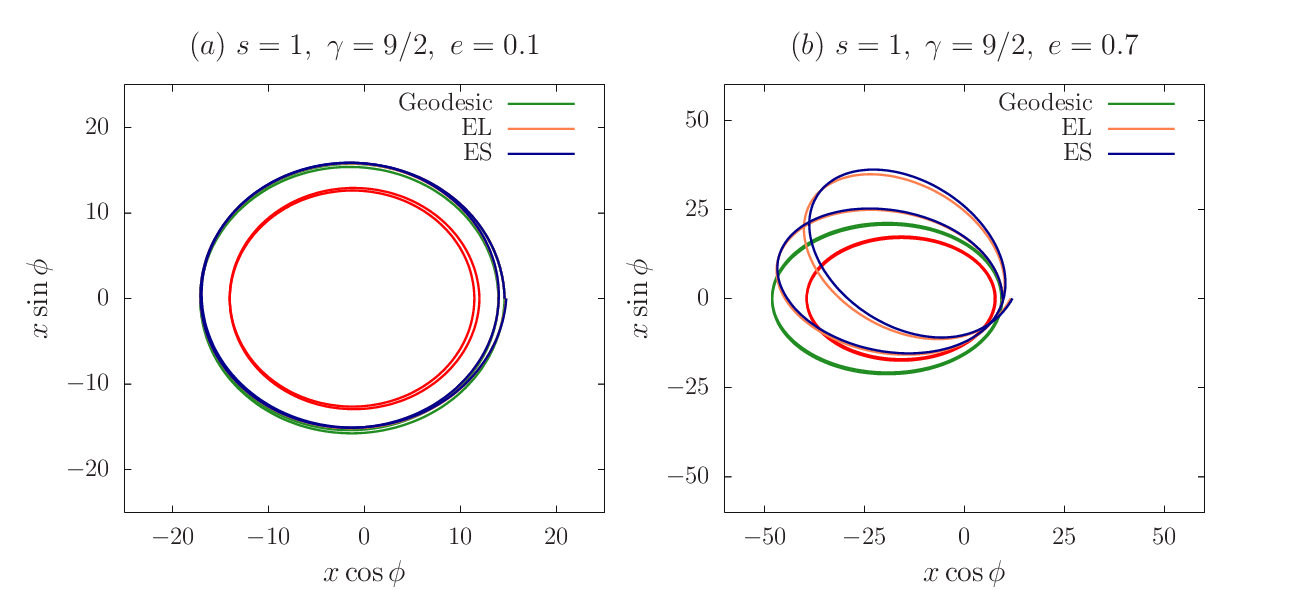}
\caption{Comparison of orbital motion for timelike test particles around renormalization group improved Schwarzschild black holes. Two approximate orbits from the EL (coral curve) and ES (dark-blue curve) methods are compared to equatorial orbits (forest-green curve), which are derived from high-precision numerical solutions for eqs.~\eqref{lageEr}--\eqref{lagelr} and \eqref{tdchi}. Parameters: $M=10^7M_{\odot}$, $\gamma=9/2$, $\lambda_{+}=1.0$, $\Omega=0.204$, $l=3.6$, $E=0.95891696$. The frequency relation
$\Upsilon_x/(2\pi)=0.2$ mHz corresponds to $\Upsilon_{\phi}/(2\pi)=0.247198$ mHz. The red curve represents the Schwarzschild case ($\gamma=\lambda_{+}=\Omega=0$).}
\label{fig:2}
\end{figure*}

We obtain the following orbital results for timelike test particles around scale-dependent Planck stars (with $s=-1$), as shown in figure~\ref{fig:1}. The figure illustrates two approximate methods for determining orbits, compared to the equatorial orbits based on geodesic equations, which are derived from high-precision numerical solutions for eqs.~\eqref{lageEr}--\eqref{lagelr} and \eqref{tdchi}. The first method is the EL method, described by eqs.~\eqref{elone}--\eqref{elthree}. The second method is the ES method, detailed in eqs.~\eqref{esone}--\eqref{esthree}.
It is worth emphasising that both the EL and ES methods are approximate. In this context, the scale-dependent Planck stars are treated as supermassive black holes with a mass of $M=10^7M_{\odot}$. The parameters used are $\gamma=9/2$ and $\lambda_{-}=-1.0$. Using eq.~\eqref{Omegafu}, the parameter $\Omega$ is calculated to be $176.392$. The energy is $E=0.99446821$, and the angular momentum is $l=8.6$. The frequency of $x$ is $\Upsilon_x/(2\pi)=0.2$ mHz, which corresponds to an approximate orbital frequency of $\Upsilon_{\phi}/(2\pi)=0.345746$ mHz. The equatorial orbits is represented by the forest-green curve, the EL method by the coral curve, and the ES method by the dark-blue curve. In subfigures $(a)$ and $(b)$, when $\gamma=\lambda_{-}=\Omega=0$, the scale-dependent Planck star reduces to a Schwarzschild black hole, with the corresponding orbital motion represented by the red curve. As seen in figure~\ref{fig:1}$(a)$, when the orbital eccentricity $e$ is small (eg., $e=0.1$), the results of the two methods are very close to each other and also closely resemble the equatorial orbits. However, as the orbital eccentricity increases ($e=0.7$), the differences between the two methods become more pronounced, as shown in figure~\ref{fig:1}$(b)$. We observe significant differences between the two methods, as well as differences compared to equatorial orbits. Clearly, the orbital motion of the timelike test particles around scale-dependent Planck star differs significantly from that around Schwarzschild black holes, whether at large or small orbital eccentricity. This indirectly suggests that both methods are not well suited for studying the orbital motion of timelike test particles around scale-dependent Planck stars at high eccentricities.

Meanwhile, figure~\ref{fig:2} shows the orbital motion of timelike test particles around renormalization group improved Schwarzschild black holes $(s=1)$ for the two approximate methods, compared with the equatorial orbits derived from geodesic equations. The mass of the renormalization group improved Schwarzschild black holes is set to $M=10^7M_{\odot}$, with parameters $\gamma=9/2$ and $\lambda_{-}=-1.0$. Using eq.~\eqref{Omegazh}, the parameter $\Omega$ is calculated to be $0.204$. The energy is $E=0.95891696$, and the angular momentum is $l=3.6$. The frequency of $x$ is $\Upsilon_x/(2\pi)=0.2$ mHz, and the corresponding approximate orbital frequency is $\Upsilon_{\phi}/(2\pi)=0.247198$ mHz. The EL is represented by the coral curve, while the ES is represented by the dark-blue curve. The forest-green curve represents the high-precision numerical result for eqs.~\eqref{lageEr}--\eqref{lagelr} and \eqref{tdchi}, which describe the equatorial orbits. Similarly, when $\gamma=\lambda_{+}=\Omega=0$, the renormalization group improved Schwarzschild black holes reduced to the Schwarzschild black holes, as depicted by the red curve in figure~\ref{fig:2}. As shown in figure~\ref{fig:2}$(a)$, for $e=0.1$, the orbits obtained by the two methods and equatorial orbits nearly overlap, with only minor differences. As seen in figure~\ref{fig:2}$(b)$, when $e=0.7$, the difference between the two methods remains small. Importantly, the orbital motion of the timelike test particles around renormalization group improved Schwarzschild black holes does not differ significantly from that around Schwarzschild black holes, whether at large or small orbital eccentricity.
To conclude, this suggests that the orbits of the timelike test particles, studied using these two methods, can help distinguish scale-dependent Planck stars from renormalization group improved Schwarzschild black holes. By contrast, the EL and ES methods are better suited for low orbital eccentricity and are more effective for analyzing renormalization group improved Schwarzschild black holes.

\section{Gravitational waves}
\label{sec:4}
The significance of gravitational wave detection lies not only in the direct capture of gravitational wave signals to confirm general relativity, but also in its broader scientific value, especially in advancing the emerging field of gravitational wave astronomy. Binary black holes, binary neutron stars, and black hole-neutron star mergers are key sources of gravitational waves. Moreover, supermassive binary black holes are among the most important and dominant sources for space-based detection of gravitational waves. A stellar-mass object orbiting a supermassive black hole is considered an EMRIs. A timelike test particle could represent the stellar-mass object, while scale-dependent Planck stars or renormalization group improved Schwarzschild black holes could serve as the supermassive black hole. The gravitational waves emitted by these EMRIs provide valuable information about the periodic orbit and the properties of the supermassive black hole.

\subsection{Energy and angular momentum fluxes of gravitational waves}
\label{Sec:4.1}
The waveform method of AK~\cite{Barack2004PRD69.082005,Babak2007PRD75.024005} is applied to timelike test particles on the equatorial plane. Based on the geodesic equations of motion, this method enables the calculation of orbital frequency and eccentricity during adiabatic evolution, which is driven by the extraction of gravitational wave energy and the angular momentum flux from the system.
The gravitational wave energy flux represents the energy carried away by gravitational waves,
while the angular momentum flux quantifies the loss of angular momentum. As gravitational waves propagate, their energy and angular momentum extraction alters the system's dynamics, driving the adiabatic orbital decay. Changes in angular momentum flux directly influence the orbital eccentricity, highlighting its critical role in the evolution of the system mediated by gravitational waves. This section focuses on calculating the energy flux and angular momentum flux of gravitational waves associated with scale-dependent Planck stars and renormalization group improved Schwarzschild black holes using the quadrupole moment approximation.

The mass moment related to the quadrupole moment is denoted as
\begin{eqnarray}
\label{sjjone}
\mathcal{M}^{ij}=\int\mu x_0^i x_0^j\mathrm{d}V,
\end{eqnarray}
where $x_0$ represents the relative position between the scale-dependent Planck stars (or renormalization group improved Schwarzschild black holes) and timelike test particles. The timelike test particles can be approximated as a point particle in the center-of-mass coordinates. Its second mass moment can be simplified as
\begin{eqnarray}
\label{sjjtwo}
\mathcal{M}^{ij}=\mu x_0^i x_0^j.
\end{eqnarray}
It is important to note that the reduced mass, $\mu=mM/(m+M)$, where $m$ and $M$ represent the masses of small object (timelike test particles) and the massive black holes (scale-dependent Planck stars or renormalization group improved Schwarzschild black holes), respectively. Under the weak-field linear approximation, we have $x_0^{i}=(x\cos{\phi},x\sin{\phi},0)$. The non-zero mass moments of the system can be calculated from the mass moment described in eq.~\eqref{sjjtwo}, denoted as:
\begin{eqnarray}
\label{qxyzone}
\mathcal{M}^{11}&=&\mu x^2\cos^2{\phi},\\
\label{qxyztwo}
\mathcal{M}^{22}&=&\mu x^2\sin^2{\phi},\\
\label{qxyzthree}
\mathcal{M}^{12}&=&\mu x^2\cos{\phi}\sin{\phi},\\
\label{qxyzfour}
\mathcal{M}^{21}&=&\mu x^2\sin{\phi}\cos{\phi}.
\end{eqnarray}
Based on the this description, the quadrupole moment tensor can be expressed as
\begin{eqnarray}
\label{sjjz}
\mathcal{Q}^{ij}=\mathcal{M}^{ij}-\frac{1}{3}\delta^{ij}\mathcal{M}_{kk}.
\end{eqnarray}
 The tensor for the non-zero quadrupole moment quantities can be described as
\begin{eqnarray}
\label{qoneone}
\mathcal{Q}^{11}&=&\mu x^2\cos^2{\phi}-\frac{1}{3}\mu x^2,\\
\label{qtwotwo}
\mathcal{Q}^{22}&=&\mu x^2\sin^2{\phi}-\frac{1}{3}\mu x^2,\\
\label{qonetwo}
\mathcal{Q}^{12}&=&\mu x^2\cos{\phi}\sin{\phi},\\
\label{qtwoone}
\mathcal{Q}^{21}&=&\mu x^2\sin{\phi}\cos{\phi},\\
\label{qthth}
\mathcal{Q}^{33}&=&-\frac{1}{3}\mu x^2,
\end{eqnarray}
with the quadrupole moment tensor given by
\begin{eqnarray}
\label{qzonghe}
\mathcal{Q}^{ij}&=&\mathcal{Q}^{11}+\mathcal{Q}^{22}+\mathcal{Q}^{12}+\mathcal{Q}^{21}+\mathcal{Q}^{33}.
\end{eqnarray}
For $v/c\ll 1$, the leading term is given by quadrupole radiation. The energy flux and angular momentum flux can be described as~\cite{Maggiore2007gwte.book,Poisson2008CQG25.209002,Thorne1980RMP52.299}:
\begin{eqnarray}
\label{nenliu}
\frac{\mathrm{d}E}{\mathrm{d}t}&=&-\frac{1}{5}\langle\dddot{\mathcal{Q}}_{ij}
\dddot{\mathcal{Q}}_{ij}\rangle,
\end{eqnarray}
and
\begin{eqnarray}
\label{jiaoliu}
\frac{\mathrm{d}l}{\mathrm{d}t}&=&-\frac{2}{5}\epsilon^{ikl}\langle
\ddot{\mathcal{Q}}_{ka}\dddot{\mathcal{Q}}_{la}\rangle,
\end{eqnarray}
where $\epsilon^{ikl}$ is the Levi-Civita symbol. The indices $i$, $k$ and $l$ refer to the spatial components, and the expression represents the flux of angular momentum carried by the gravitational waves. The $\epsilon^{ikl}$ symbol ensures that the flux is described in terms of the cross product of the components of the second time derivatives of the quadrupole moment tensor, leading to the angular momentum vector. The multipole expansion assumes that the orbital motion is non-relativistic, meaning that the dynamics of the orbit's radiation are governed by Newtonian mechanics, at least at the lowest order terms.

When derived over time and combined with the harmonic law, the expression for the energy flux of gravitational waves can be specifically written as:
\begin{eqnarray}
\label{nlbdzong}
\frac{\mathrm{d}E}{\mathrm{d}t}&=&\frac{\mathrm{d}E}{\mathrm{d}t}\Big|_{\mathrm{Schw}}+\frac{\mathrm{d}E}{\mathrm{d}t}\Big|_{\mathrm{s\Omega}},
\end{eqnarray}
where
\begin{eqnarray}
\label{nlbdone}
\frac{\mathrm{d}E}{\mathrm{d}t}\Big|_{\mathrm{Schw}}&=&-\frac{(1-e^2)^{3/2}}{15p^5}(96+292e^2+37e^4)\mu^2+\mathcal{O}(p^{-6}),
\end{eqnarray}
and
\begin{eqnarray}
\label{nlbdtwo}
\frac{\mathrm{d}E}{\mathrm{d}t}\Big|_{\mathrm{s\Omega}}
&=&\frac{(1-e^2)^{3/2}}{30p^6}(37e^6-111e^4-1164e^2 -288)\mu^2
+\frac{(1-e^2)^{3/2}}{120p^7}(111e^8+392e^6
\nonumber \\
&&-6517e^4-19668e^2-4320)\mu^2+\frac{(1-e^2)^{3/2}}{30p^7}s\Omega(-37e^6+111e^4+1164e^2
\nonumber \\
&&+288)\mu^2+\mathcal{O}(p^{-8}).
\end{eqnarray}
The angular momentum flux for gravitational waves is given by
\begin{eqnarray}
\label{jlbdzong}
\frac{\mathrm{d}l}{\mathrm{d}t}&=&\frac{\mathrm{d}l}{\mathrm{d}t}\Big|_{\mathrm{Schw}}+\frac{\mathrm{d}l}{\mathrm{d}t}\Big|_{\mathrm{s\Omega}},
\end{eqnarray}
where
\begin{eqnarray}
\label{jlbdone}
\frac{\mathrm{d}l}{\mathrm{d}t}\Big|_{\mathrm{Schw}}&=&-\frac{4(1-e^2)^{3/2}}{5p^{7/2}}(8+7e^2)\mu^2+\mathcal{O}(p^{-9/2}),
\end{eqnarray}
and
\begin{eqnarray}
\label{jlbdtwo}
\frac{\mathrm{d}l}{\mathrm{d}t}\Big|_{\mathrm{s\Omega}}&=&\frac{(1-e^2)^{3/2}}{5p^{9/2}}(9e^4-66e^2-48)\mu^2
+\frac{(1-e^2)^{3/2}}{10p^{11/2}}(16e^6-42e^4-643e^2-360)\mu^2
\nonumber \\
&&+\frac{(1-e^2)^{3/2}}{5p^{11/2}}s\Omega(-9e^4+66e^2+48)\mu^2+\mathcal{O}(p^{-13/2}).
\end{eqnarray}
Eqs.~\eqref{nlbdzong}--\eqref{nlbdtwo} and \eqref{jlbdzong}--\eqref{jlbdtwo} describe the fluxes of energy and angular momentum, such as timelike test particles around scale-dependent Planck stars or renormalization group improved Schwarzschild black holes. These equations are crucial for understanding the evolution of these objects and their interactions. They also highlight the critical role of energy and angular momentum dissipation due to gravitational radiation in the analysis of astrophysical systems.

\subsection{Numerical computation of gravitational waveforms}
\label{Sec:4.2}
Based on Eqs.~\eqref{nlbdzong}--\eqref{nlbdtwo} and \eqref{jlbdzong}--\eqref{jlbdtwo}, we determined the gravitational waveforms, incorporating the effects of gravitational radiation. In a strong gravitational field, the bound orbits of timelike test particles around a massive object will exhibit periodic or quasi-periodic behavior~\cite{Huang2025PRD111.084038,Deng2021PDU31.100745,Deng2023EPJC83.311}. According to the classification method for particle periodic orbits proposed in refs.~\cite{Levin2008PRD77.103005,Levin2009PRD79.043016}, these orbits are characterized by three integers $(z,w,v)$, which define a rational number $q$. The integers are defined as follows: $q=w+v/z$, where $z$ represents the number of ``zooms" of the periodic orbit, $w$ is the number of whirl around the center, and $v$ describes the vertex behavior of the periodic orbit. 

\begin{figure*}
\centering
\includegraphics[width=1.04\textwidth]{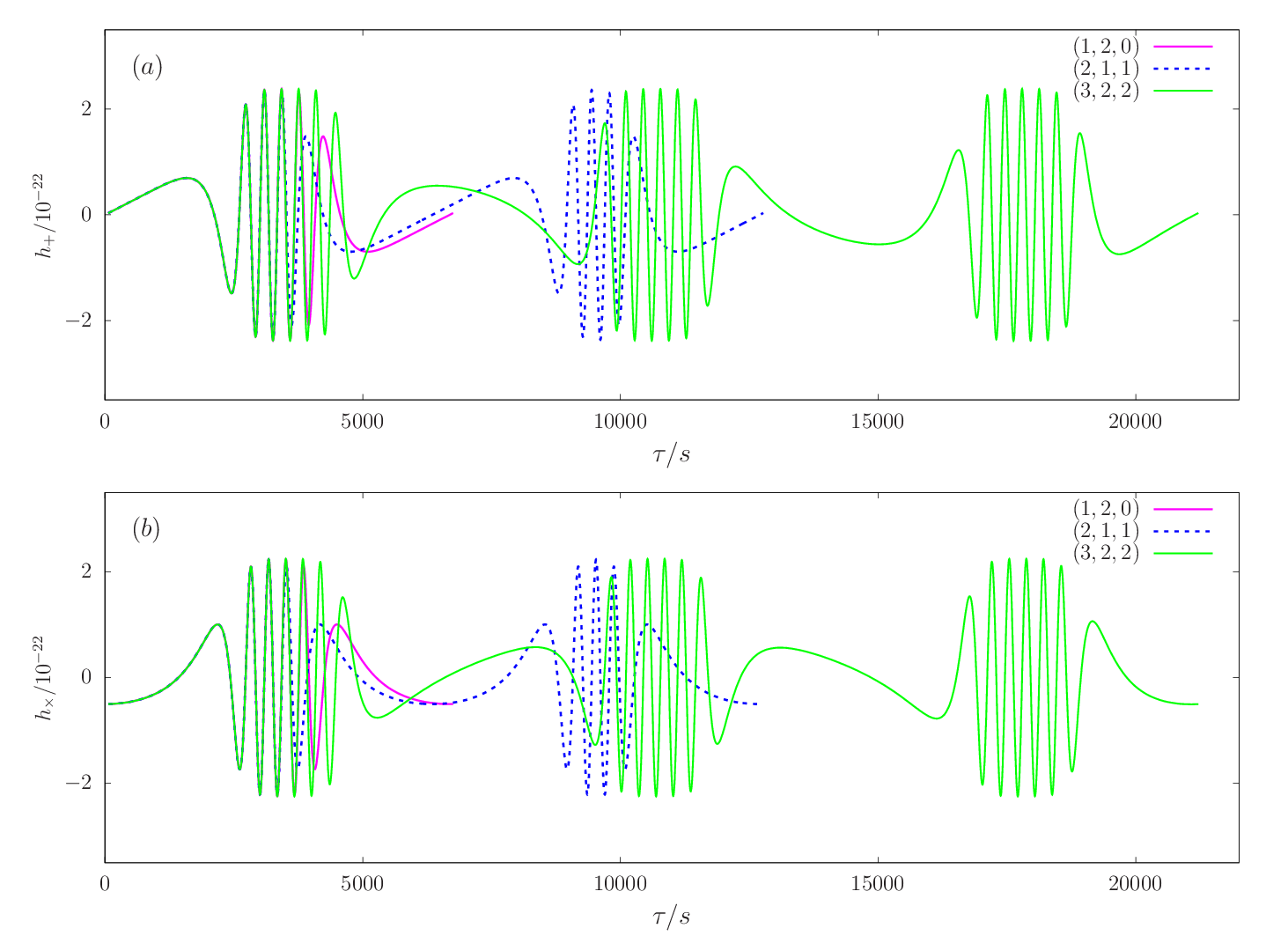}
\caption{Gravitational waveforms from different periodic orbits of timelike test particles ($m=10M_{\odot}$) around scale-dependent Planck stars ($M=10^7M_{\odot}$) with parameters $\gamma=9/2$, $\lambda_{-}=-1.0$, and initial angular momentum $l=8.6$. The magenta, blue, and green curves represent periodic orbits $(1,2,0)$, $(2,1,1)$, and $(3,2,2)$, with corresponding initial energies of $0.99450576$, $0.99450395$, and $0.99450586$, respectively.}
\label{fig:3}
\end{figure*}

The numerical Kludge waveform model has proven effective for investigating gravitational waveforms from EMRIs~\cite{Babak2007PRD75.024005}. Specifically, the transverse traceless tensor polarizations $h_{+}$ and $h_{\times}$ can be described as~\cite{Poisson2014Book}:
\begin{eqnarray}
\label{hplus}
h_{+}&=&-\frac{2\eta}{c^2D_{L}}\frac{GM}{x}(1+{\cos}^2{\iota})\cos{(2\phi+2\zeta)},\\
\label{hcross}
h_{\times}&=&-\frac{4\eta}{c^2D_{L}}\frac{GM}{x}{\cos}{\iota}\sin{(2\phi+2\zeta)}.
\end{eqnarray}
Where $\phi$ denote the phase angle determined by eq.~\eqref{lagelr}, $\eta=mM/(m+M)^2$. We consider a small object with mass $m=10M_\odot$ along different periodic orbits around a supermassive (such as scale-dependent Planck stars or renormalization group improved Schwarzschild black holes) with mass $M=10^7M_\odot$, located at a luminosity distance of $D_{L}=200~\mathrm{Mpc}$. The inclination angle is set to $\iota=\pi/4$, and the longitude of pericenter is set to $\zeta=\pi/4$. We compute the two transverse traceless tensor polarizations described by eq.~\eqref{hplus} and eq.~\eqref{hcross} for the gravitational waveforms generated by the different periodic orbits of timelike test particles ($m=10M_{\odot}$) around scale-dependent Planck stars, with parameters $\gamma=9/2$ and $\lambda_{-}=-1.0$. The numerical results are plotted in figure~\ref{fig:3}, where a time-domain approach is employed to analyze the gravitational waveforms, while the initial angular momentum $l=8.6$ for eq.~\eqref{jlbdzong}, and periodic orbits $(1,2,0)$, $(2,1,1)$, and $(3,2,2)$ are considered, with corresponding initial energies of $0.99450576$, $0.99450395$, and $0.99450586$, respectively, for eq.~\eqref{nlbdzong}. The transverse traceless tensor polarizations are shown as a function of the proper times $\tau(\mathrm{s})$. The megenta, blue, and green curves correspond to the periodic orbits $(1,2,0)$, $(2,1,1)$, and $(3,2,2)$, respectively. Both polarizations, $h_{+}$ and $h_{\times}$, clearly demonstrate the zoom-whirl behavior of the corresponding periodic orbits. The number of zooms in the periodic orbit matches the static phase of the gravitational wave, while the number of whirls corresponds to the louder glitches in the signal.

\begin{figure*}
\centering
\includegraphics[width=1.04\textwidth]{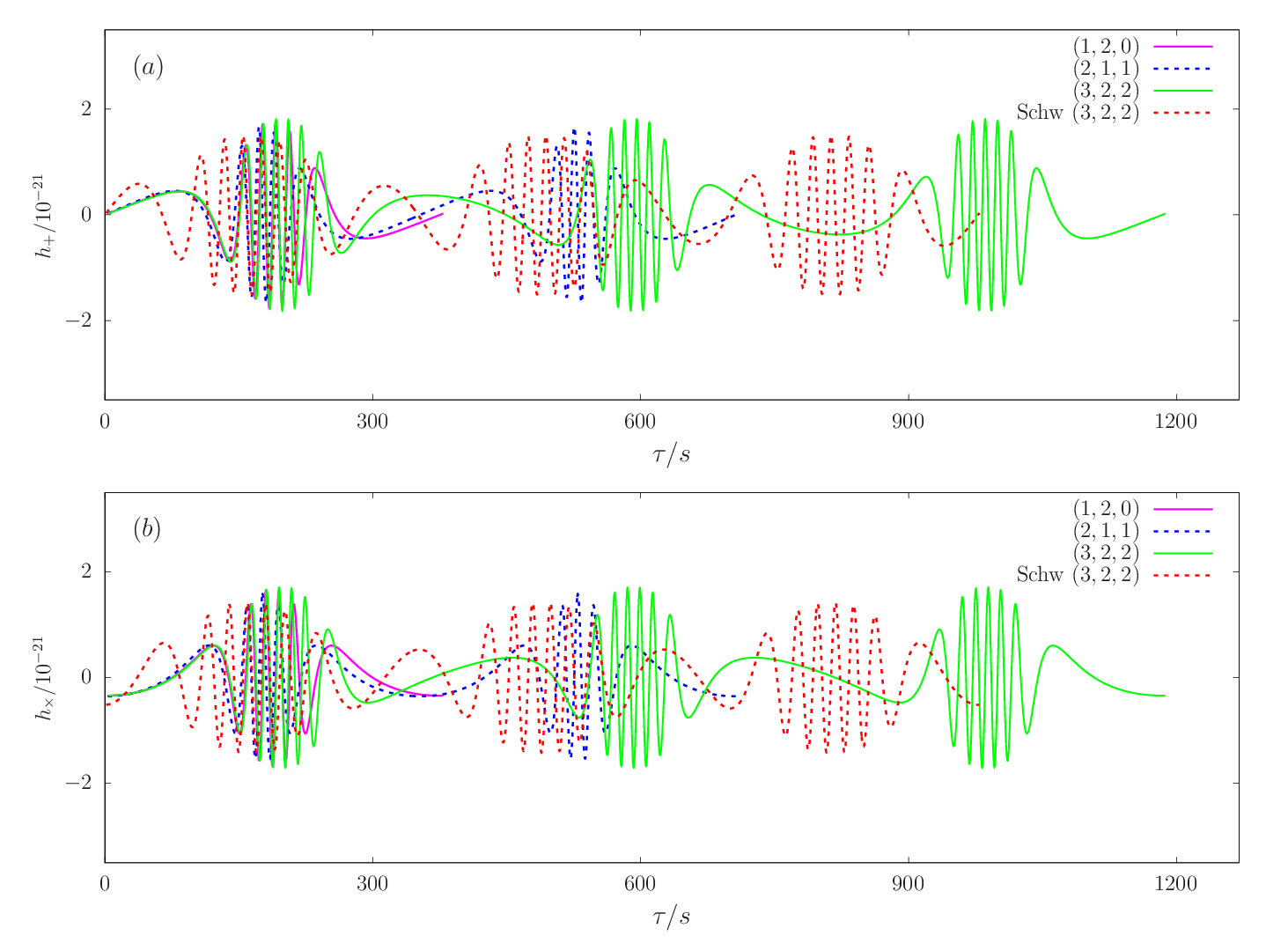}
\caption{Gravitational waveforms from different periodic orbits of timelike test particles ($m=10M_{\odot}$) around renormalization group improved Schwarzschild black holes ($M=10^7M_{\odot}$) with parameters $\gamma=9/2$, $\lambda_{+}=1.0$, and initial angular momentum $l=3.6$. The magenta, blue, and green curves correspond to periodic orbits $(1,2,0)$, $(2,1,1)$, and $(3,2,2)$, respectively, with initial energies of $0.96367919$, $0.96298541$, and $0.96381768$, respectively. In panels $(a)$ and $(b)$, when the parameter $\gamma=\lambda_{+}=\Omega=0$, the system reduces to the Schwarzschild black holes. The red curve represents the periodic orbit $(3,2,2)$ of timelike test particles around Schwarzschild black holes, with a corresponding initial energy $E=0.95471456$ and initial angular momentum $l=3.6$.}
\label{fig:4}
\end{figure*}

Furthermore, figure~\ref{fig:4} depicts the gravitational waveforms generated by the periodic orbits $(1,2,0)$, $(2,1,1)$, and $(3,2,2)$ of timelike test particles with mass $m=10M_{\odot}$ around renormalization group improved Schwarzschild black holes ($M=10^7M_{\odot}$). The relevant parameters are chosen as $\gamma=9/2$ and $\lambda_{+}=1.0$. 
The initial angular momentum is set to $l=3.6$ for eq.~\eqref{jlbdzong}. Periodic orbits $(1,2,0)$, $(2,1,1)$, and $(3,2,2)$ are considered, with corresponding initial energies of $0.96367919$, $0.96298541$, and $0.96381768$, respectively, for eq.~\eqref{nlbdzong}. when the parameter $\gamma=\lambda_{+}=\Omega=0$, the system reduces to the Schwarzschild black holes. This figure also presents the gravitational waveforms from the periodic orbits $(3,2,2)$ of a timelike test particle around Schwarzschild black holes. The initial energy is $E=0.95471456$ for eq.~\eqref{nlbdzong}. By comparing figure~\ref{fig:3} with figure~\ref{fig:4}, we observe that the periodic orbits of timelike test particles around renormalization group improved Schwarzschild black holes also reflect their number of static phases, which correspond to the number of zooms. Nonetheless, the magnitude of the corresponding gravitational waveform amplitude is on the order of $10^{-21}$, while for scale-dependent Planck stars, it is on the order of $10^{-22}$. Moreover, the time required for scale-dependent Planck stars to complete a single period of motion is significantly longer than that for renormalization group improved Schwarzschild black holes. This demonstrates that the gravitational waveforms of these periodic orbits can help distinguish scale-dependent Planck stars from renormalization group improved Schwarzschild black holes.

\begin{figure*}
\centering
\includegraphics[width=1.06\textwidth]{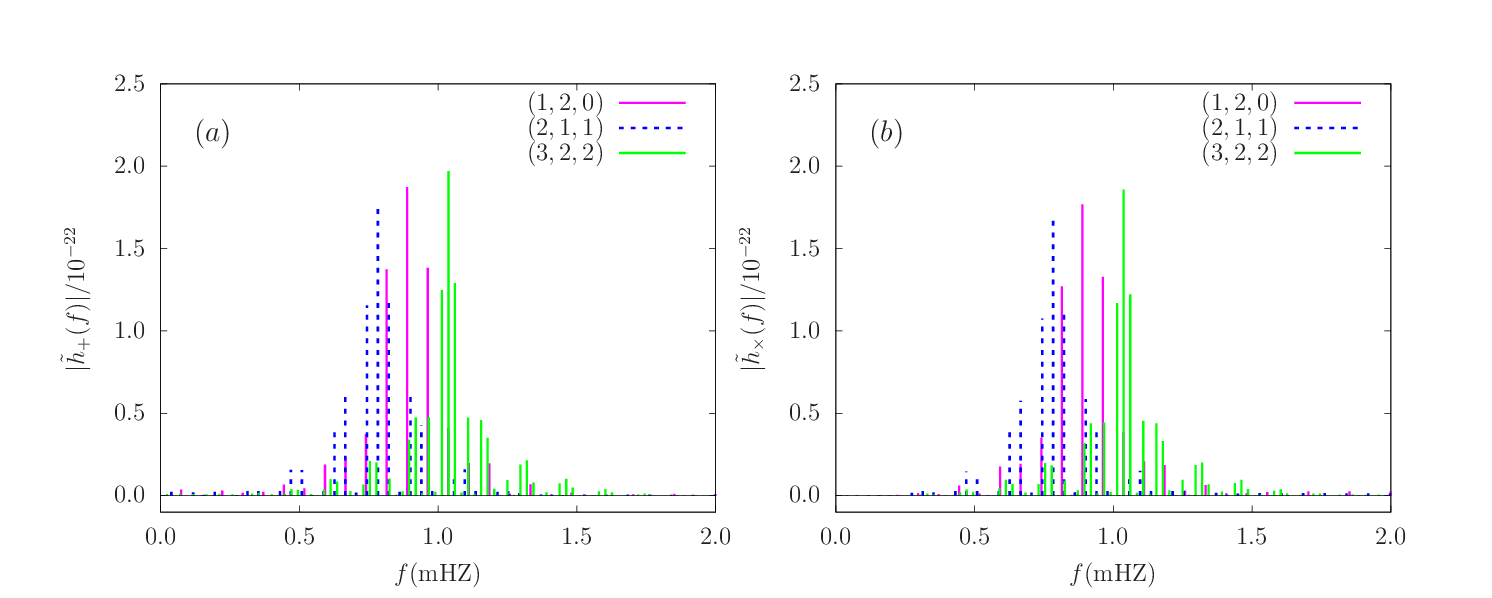}
\caption{The absolute values of the discrete Fourier transforms of gravitational waveforms from periodic orbits $(1,2,0)$, $(2,1,1)$, and $(3,2,2)$ of timelike test particles ($m=10M_{\odot}$) around scale-dependent Planck stars with mass $M=10^7M_{\odot}$, calculated using parameters $\gamma=9/2$, $\lambda_{-}=-1.0$, and initial angular momentum $l=8.6$.}
\label{fig:5}
\end{figure*}

\begin{figure*}
\centering
\includegraphics[width=1.06\textwidth]{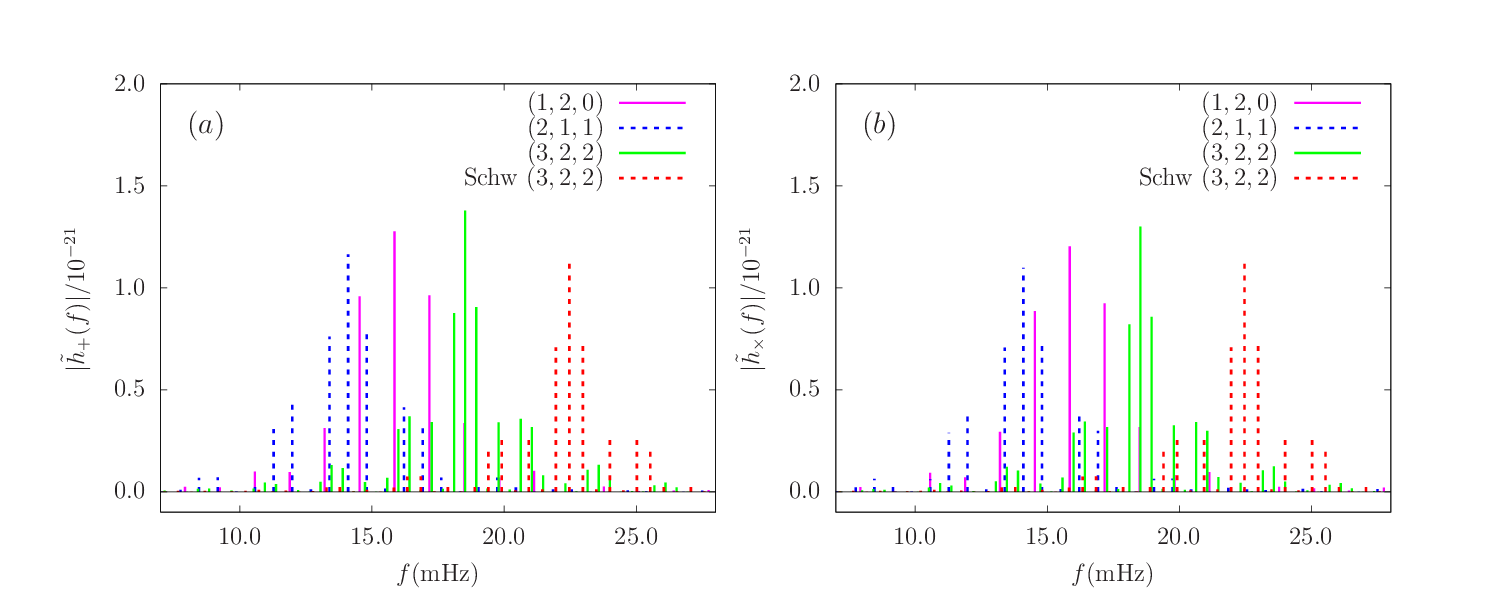}
\caption{The absolute values of the discrete Fourier transforms of gravitational waveforms from different periodic orbits of timelike test particles with $m=10M_{\odot}$ around renormalization group improved Schwarzschild black holes ($M=10^7M_{\odot}$, $\gamma=9/2$, $\lambda_{+}=1.0$), with initial angular momentum $l=3.6$. In panels $(a)$ and $(b)$, when the parameter $\gamma=\lambda_{+}=\Omega=0$, the system reduces to Schwarzschild black holes, with periodic orbits $(3,2,2)$ shown by the red curves.}
\label{fig:6}
\end{figure*}

By applying a discrete Fourier transform to the time-domain gravitational waveforms, we obtain the corresponding frequency spectra shown in figure~\ref{fig:3} and figure~\ref{fig:4}. This demonstrates the use of the frequency-domain to study the periodic orbits of timelike test particles around supermassive black holes. For example, we plot the absolute values of the frequency spectra $\tilde{h}_{+}(f)$ and $\tilde{h}_{\times}(f)$ corresponding to the timelike test particles periodic orbits $(1,2,0)$, $(2,1,1)$, and $(3,2,2)$ around scale-dependent Planck stars and those around renormalization group improved Schwarzschild black holes. These are shown in figure~\ref{fig:5} and figure~\ref{fig:6}. In contrast, the frequency-domain spectra of the two spacetimes exhibit distinct differences. The characteristic frequency range of scale-dependent Planck stars is concentrated between $0$ mHz and $2$ mHz, whereas for renormalization group improved Schwarzschild black holes, it lies between $7$ mHz and $28$ mHz. The gravitational wave amplitudes in the frequency-domain exhibit the same distinctions observed in the time-domain. Furthermore, the red curve in figure~\ref{fig:6} represents the case of Schwarzschild black holes, which closely resemble the physical properties of renormalization group improved Schwarzschild black holes, while showing a greater divergence from those of scale-dependent Planck stars. In summary, the gravitational waves in the frequency-domain can similarly distinguish scale-dependent Planck stars from renormalization group improved Schwarzschild black hole. Based on this, we can infer that the characteristic frequencies of the gravitational waveforms from EMRIs with periodic orbits generally fall within the millihertz range, which is detectable by space-based detectors. Next, we will examine the corresponding characteristic strains in these two spacetimes and compare them with those of ground-based and space-based detectors. This comparison will provide a clearer indication of the detectability of  scale-dependent Planck stars and renormalization group improved Schwarzschild black hole using both ground-based and space-based detectors.

\subsection{Gravitational wave sensitivity curves}
In gravitational wave astronomy, the detection capability of detectors is quantitatively characterized by the sensitivity curve, also known as the noise power spectral density curve.
This curve delineates the equivalent noise level of the detector across specific Fourier frequency ranges, with the vertical axis typically expressed in logarithmic coordinates of amplitude spectral density or characteristic strain ($h_{\mathrm{c}}(f)$).
When assessing the observability of potential gravitational wave sources, a frequency-domain comparison is performed between the source's gravitational wave emission spectrum (i.e., characteristic strain spectrum) and the detector's noise curve. If the characteristic strain noise amplitude of the target source consistently exceeds the detector's noise floor ($h_{\mathrm{source}}>h_{\mathrm{n}}(f)$) within specific frequency intervals, the signal can be extracted from instrumental noise using advanced signal processing techniques, such as matched filtering, enabling successful detection (with a signal-to-noise ratio exceeding the detection threshold).
Conversely, if the source spectrum lies entirely below the noise curve, the signal remains buried within the detector's intrinsic noise and fails to meet the statistical significance requirements for detection. This frequency-domain comparison methodology serves as the fundamental criterion for evaluating the feasibility of gravitational wave source detection. The observational feasibility of gravitational waveforms from periodic orbits of timelike test particles around scale-dependent Planck stars or renormalization group improved Schwarzschild black holes is assessed by calculating the relevant dimensionless characteristic strain~\cite{Finn2000PRD62.l24021}:
\begin{eqnarray}
\label{tzyb}
h_{\mathrm{c}}(f)&=&2f\sqrt{\big|\tilde{h}_{+}(f)\big|^2+\big|\tilde{h}_{\times}(f)\big|^2}.
\end{eqnarray}
This is then compared with the sensitivity curves of LISA~\cite{Robson2019CQG36.105011}, eLISA~\cite{Amaro-Seoane2012CQG29.124016}, TianQin~\cite{Luo2016CQG33.035010}, BBO~\cite{Cutler2006PRD73.042001}, DECIGO~\cite{DECIGO2021PTEP2021.05A105}, EPTA~\cite{Antoniadis2023A&A678.A50}, IPTA~\cite{Hobbs2010CQG27.084013}, SKA \cite{Braun2015SKA174}, LIGO~\cite{LIGO2017CQG34.044001}, aLIGO~\cite{LIGO2021PRX11.021053}, and LIGO~$\mathrm{A^{+}}$~\cite{Tso2022Aplus}.

\begin{figure*}
\centering
\includegraphics[width=1.0\textwidth]{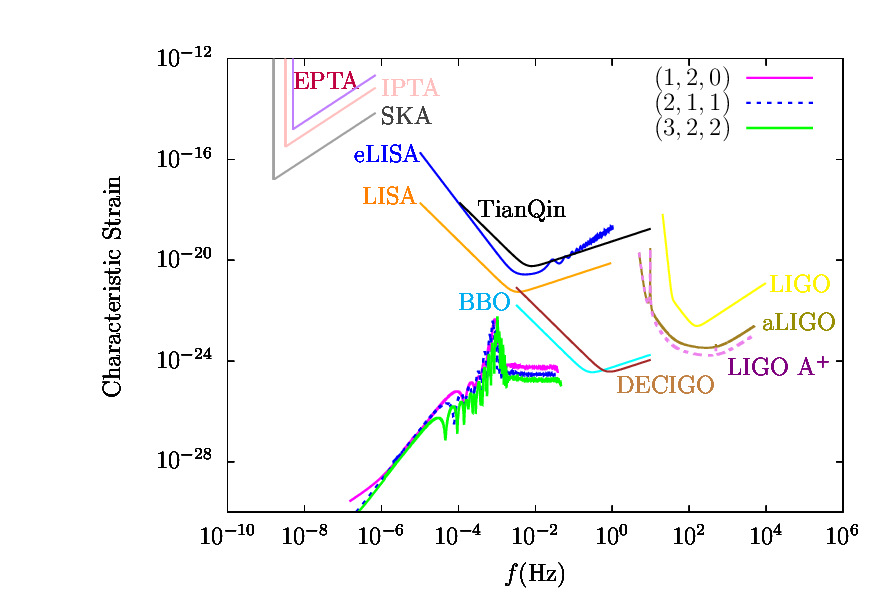}
\caption{Comparison of the characteristic strain of gravitational waves in different periodic orbits (figure~\ref{fig:5}) compared to the sensitivity curves of LISA (orange), eLISA (blue), TianQin (black), BBO (cyan), DECIGO (brown), EPTA (purple), IPTA (pink), SKA (dark-gray), LIGO (yellow), aLIGO (olive), and LIGO~$\mathrm{A^{+}}$ (violet).}
\label{fig:7}
\end{figure*}

Based on the analysis and description above, we plot the characteristic strain $h_{\mathrm{c}}(f)$ of the gravitational waves from timelike test particles around scale-dependent Planck stars or renormalization group improved Schwarzschild black holes.
We then overlay the sensitivity curves of all the detectors mentioned above on the same plot, allowing a comparison of their sensitivities with the characteristic strain curve.
figure~\ref{fig:7} shows that the characteristic strain for scale-dependent Planck stars, with varying values of $(w,v,z)$, remains below the sensitivity curves of the aforementioned detectors. This indirectly suggests that these detectors are unable to detect the periodic orbit gravitational wave signals from timelike test particles around scale-dependent Planck stars. Additionally, figure~\ref{fig:7} illustrates the frequency ranges of sensitivity for different detectors: LISA, eLISA, and TianQin, as space-based detectors, are sensitive to low-frequency gravitational waves in the range from $10^{-4}$ Hz to $1$ Hz.
BBO and DECIGO, future space-based detectors, cover a frequency range form roughly $10^{-2}$ Hz to $10$ Hz. EPTA, IPTA, and SKA, as pulsar timing arrays, focus on low-frequency gravitational waves with sensitivity between $10^{-9}$ Hz and $10^{-6}$ Hz. LIGO, aLIGO, and LIGO~$\mathrm{A^{+}}$, as ground-based detectors, are sensitive to frequencies between $10$ Hz and a few kHz.

\begin{figure*}
\centering
\includegraphics[width=1.0\textwidth]{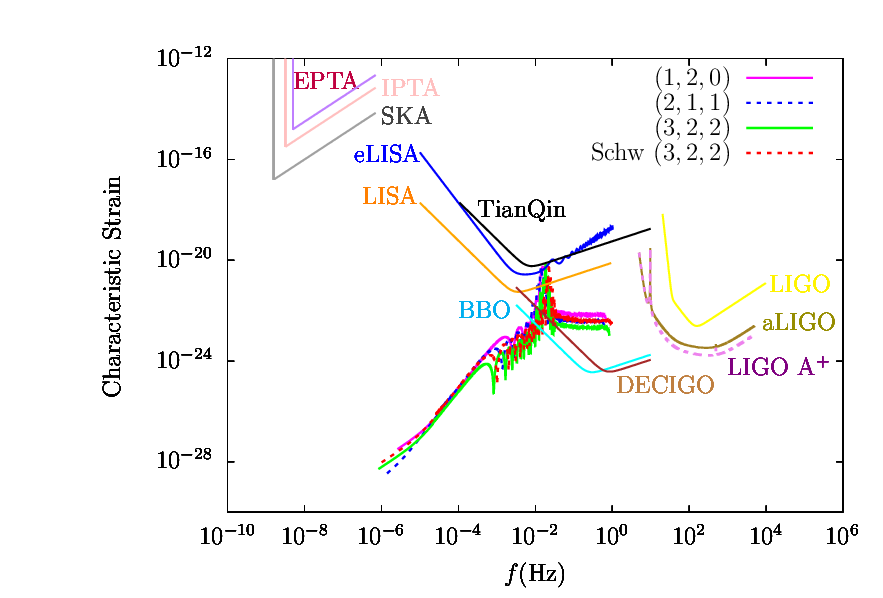}
\caption{Comparison of the characteristic strain of gravitational waves in different periodic orbits (figure~\ref{fig:6}) compared to the sensitivity curves of LISA (orange), eLISA (blue), TianQin (black), BBO (cyan), DECIGO (brown), EPTA (purple), IPTA (pink), SKA (dark-gray), LIGO (yellow), aLIGO (olive), and LIGO~$\mathrm{A^{+}}$ (violet).}
\label{fig:8}
\end{figure*}

To better distinguish scale-dependent Planck stars from renormalization group improved Schwarzschild black holes, we examine the characteristic strain of the gravitational waves from timelike test particles around renormalization group improved Schwarzschild black holes with different $(w,v,z)$, as shown in figure~\ref{fig:8}. It also presents the characteristic strains of the periodic orbits $(3,2,2)$ of timelike test particles around Schwarzschild black holes. The most striking aspect of the figure is that the corresponding characteristic strains exceed the sensitivity curves of LISA, BBO, DECIGO. This visual comparison shows that these three detectors are the most capable of detecting the gravitational wave signals from periodic orbits of timelike test particles around renormalization group improved Schwarzschild black holes. The same is true for gravitational wave signals from periodic orbits of timelike test particles around Schwarzschild black holes. Figure~\ref{fig:7} and figure~\ref{fig:8} reveal a significant difference between scale-dependent Planck stars and renormalization group improved Schwarzschild black holes, providing further observational validation of their detectability. This adequately demonstrates that
gravitational waves can distinguish scale-dependent Planck stars from renormalization group improved Schwarzschild black holes.

\section{Conclusions and discussions}
\label{sec:5}
In order to distinguish scale-dependent Planck stars from renormalization group improved Schwarzschild black holes, we analyze the orbital evolution of timelike test particles around these spacetimes, along with their gravitational waveforms. Astrophysically, this exploration reveals the physical properties of timelike test particles near supermassive black holes. We calculate the equations of motion for EMRIs and the orbital frequencies (radial frequency, azimuthal frequency, and precession frequency) on the equatorial plane in the weak-field limit. Notably, when $\gamma=\Omega=0$, the precession frequency reduces to the classical Schwarzschild case.
To obtain the orbital evolution, we use the fundamental frequencies $\nu_{r}$ and $\nu_{\phi}$, adopting two approximate methods--the EL and ES methods--and compare them with equatorial orbits based on geodesic equations. We analyze the feasibility of these methods for studying scale-dependent Planck stars and renormalization group improved Schwarzschild black holes. The results indicate that the difference between the two methods is more pronounced in scale-dependent Planck stars. In other words, these approximations are better suited for studying renormalization group improved Schwarzschild black holes. Meanwhile, we find that these two spacetimes can be distinguished based on their orbital evolution.

Based on the quadrupole moment, we calculate the energy flux and angular momentum flux due to gravitational wave emission. To consider gravitational radiation effects in EMRIs, we investigate the gravitational waves generated by periodic orbits of timelike test particles ($m=10M_\odot$) around scale-dependent Planck stars or renormalization group improved Schwarzschild black holes with $M=10^7M_\odot$. These orbits include $(1,2,0)$, $(2,1,1)$ and $(3,2,2)$. We explore the corresponding gravitational waveforms in the time-domain and numerically calculate the associated energy and orbital angular momentum. When the parameters $\gamma=\Omega=0$, the system reduces to Schwarzschild black holes. We present the gravitational waveforms form periodic orbit $(3,2,2)$ of timelike test particles around Schwarzschild black holes. We found that the magnitude of the corresponding gravitational waveforms is on the order of $10^{-21}$ for renormalization group improved Schwarzschild black holes. For scale-dependent Planck stars, the amplitude is one order of magnitude lower, around $10^{-22}$. The corresponding frequency spectra are obtained by performing discrete Fourier transforms on the time-domain gravitational waveforms. The results show that the magnitudes of the gravitational waveforms amplitudes for timelike test particles around scale-dependent Planck stars and those around renormalization group improved Schwarzschild black holes, in both the frequency and time domains, are distinct. By contrast, the key difference between the two spacetimes is that the frequency spectrum variations of the former are concentrated in the frequency range from $0$ mHz to $2$ mHz, as shown in figure~\ref{fig:5}, while those of the latter are concentrated between $7$ mHz to $28$ mHz, as shown in figure~\ref{fig:6}. Our results confirm that gravitational waveforms, both in the time-domain and the frequency-domain, can distinguish scale-dependent Planck stars from renormalization group improved Schwarzschild black holes.

According to the frequency spectra obtained from the discrete Fourier transform, we observe that the frequencies of the two spacetimes are on the order of millihertz. Our analysis suggests that the characteristic frequencies of gravitational radiation from EMRIs in periodic orbits typically lie within the sensitive frequency bands of planned space-based gravitational wave observatories. Furthermore, we calculated the corresponding characteristic strains of gravitational waves for periodic orbits $(w,v,z)$ and compared them with the gravitational wave sensitivity curves of detectors such as LISA, eLISA, TianQin, BBO, DECIGO, EPTA, IPTA, SKA, LIGO, aLIGO, and LIGO~$\mathrm{A^{+}}$. By making this comparison visually clear, we can determine which gravitational wave detectors are most capable of detecting signals from timelike test particles around scale-dependent Planck stars or renormalization group improved Schwarzschild black holes, based on the specific frequencies and characteristic strains of the gravitational waves. Our analysis reveals that detectors like LISA, BBO, DECIGO can detect signals from timelike test particles around renormalization group improved Schwarzschild black holes, but signals from scale-dependent Planck stars remain undetected. This finding suggests a distinct difference between the two spacetimes, and future space-based detectors are likely to detect gravitational waves emitted by EMRIs with periodic orbits. In fact, it has also been observationally verified that gravitational waves can distinguish scale-dependent Planck stars from renormalization group improved Schwarzschild black holes.

It is crucial to emphasize that, in this work, we adopted a rudimentary modeling framework when investigating gravitational waves emitted by EMRIs. Future studies should focus on more complex scenarios to develop the final EMRIs signal processing pipeline. In addition, we analysis of gravitational waves in periodic orbits, with numerical calculations performed using the AK method. Future research will incorporate Markov Chain Monte Carlo simulations and focus on the motion of timelike test particles around a spinning black hole. These challenges will be systematically addressed in our subsequent investigations.


\acknowledgments
We would like to thank Hou-Yu Lin for discussions on gravitational wave sensitivity curves. This work is supported by the National Natural Science Foundation of China, Grant Nos. 11903022 and 12163003. The authors also appreciate the support of the Purple Mountain Observatory, Chinese Academy of Sciences.









\bibliographystyle{JHEP}
\bibliography{biblio}

\providecommand{\href}[2]{#2}\begingroup\raggedright\begin{thebibliography}{10}

\bibitem{Einstein1915SPAW844}
A.~{Einstein}, \emph{{Die Feldgleichungen der Gravitation}},
  {\emph{Sitzungsberichte der K\&ouml;niglich Preussischen Akademie der
  Wissenschaften} (1915) 844}.

\bibitem{Holberg2010JHA41.41}
J.B.~{Holberg}, \emph{{Sirius B and the Measurement of the Gravitational
  Redshift}}, \href{https://doi.org/10.1177/002182861004100102}{\emph{Journal
  for the History of Astronomy} {\bfseries 41} (2010) 41}.

\bibitem{Hulse1975ApJ195.L51}
R.A.~{Hulse} and J.H.~{Taylor}, \emph{{Discovery of a pulsar in a binary
  system.}}, \href{https://doi.org/10.1086/181708}{\emph{\apjl} {\bfseries 195}
  (1975) L51}.

\bibitem{Clark1977ApJ215.311}
J.P.A.~{Clark} and D.M.~{Eardley}, \emph{{Evolution of close neutron star
  binaries.}}, \href{https://doi.org/10.1086/155360}{\emph{\apj} {\bfseries
  215} (1977) 311}.

\bibitem{Damour1981PLA87.81}
T.~{Damour} and N.~{Deruelle}, \emph{{Radiation reaction and angular momentum
  loss in small angle gravitational scattering}},
  \href{https://doi.org/10.1016/0375-9601(81)90567-3}{\emph{Physics Letters A}
  {\bfseries 87} (1981) 81}.

\bibitem{Damour1985AIHPA43.107}
T.~{Damour} and N.~{Deruelle}, \emph{{General relativistic celestial mechanics
  of binary systems. I. The post-Newtonian motion.}}, {\emph{Annales de
  L'Institut Henri Poincare Section (A) Physique Theorique} {\bfseries 43}
  (1985) 107}.

\bibitem{Perera2019MNRAS490.4666}
B.B.P.~{Perera}, \emph{{The International Pulsar Timing Array: second data
  release}}, \href{https://doi.org/10.1093/mnras/stz2857}{\emph{Monthly Notices
  of the Royal Astronomical Society} {\bfseries 490} (2019) 4666}
  [\href{https://arxiv.org/abs/1909.04534}{{\ttfamily 1909.04534}}].

\bibitem{Moore2015CQG32.055004}
C.J.~{Moore}, S.R.~{Taylor} and J.R.~{Gair}, \emph{{Estimating the sensitivity
  of pulsar timing arrays}},
  \href{https://doi.org/10.1088/0264-9381/32/5/055004}{\emph{Classical and
  Quantum Gravity} {\bfseries 32} (2015) 055004}
  [\href{https://arxiv.org/abs/1406.5199}{{\ttfamily 1406.5199}}].

\bibitem{LIGO2016PRL116.061102}
{LIGO Scientific Collaboration and Virgo Collaboration}, \emph{{Observation of
  Gravitational Waves from a Binary Black Hole Merger}},
  \href{https://doi.org/10.1103/PhysRevLett.116.061102}{\emph{\prl} {\bfseries
  116} (2016) 061102} [\href{https://arxiv.org/abs/1602.03837}{{\ttfamily
  1602.03837}}].

\bibitem{LIGO2016PRL116.241103}
{LIGO Scientific Collaboration and Virgo Collaboration}, \emph{{GW151226:
  Observation of Gravitational Waves from a 22-Solar-Mass Binary Black Hole
  Coalescence}},
  \href{https://doi.org/10.1103/PhysRevLett.116.241103}{\emph{\prl} {\bfseries
  116} (2016) 241103} [\href{https://arxiv.org/abs/1606.04855}{{\ttfamily
  1606.04855}}].

\bibitem{LIGO2017PRL118.221101}
{LIGO Scientific Collaboration and Virgo Collaboration}, \emph{{GW170104:
  Observation of a 50-Solar-Mass Binary Black Hole Coalescence at Redshift
  0.2}}, \href{https://doi.org/10.1103/PhysRevLett.118.221101}{\emph{\prl}
  {\bfseries 118} (2017) 221101}
  [\href{https://arxiv.org/abs/1706.01812}{{\ttfamily 1706.01812}}].

\bibitem{LIGO2017ApJ851.L35}
{LIGO Scientific Collaboration and Virgo Collaboration}, \emph{{GW170608:
  Observation of a 19 Solar-mass Binary Black Hole Coalescence}},
  \href{https://doi.org/10.3847/2041-8213/aa9f0c}{\emph{\apjl} {\bfseries 851}
  (2017) L35} [\href{https://arxiv.org/abs/1711.05578}{{\ttfamily
  1711.05578}}].

\bibitem{LIGO2017PRL119.141101}
{LIGO Scientific Collaboration and Virgo Collaboration}, \emph{{GW170814: A
  Three-Detector Observation of Gravitational Waves from a Binary Black Hole
  Coalescence}},
  \href{https://doi.org/10.1103/PhysRevLett.119.141101}{\emph{\prl} {\bfseries
  119} (2017) 141101} [\href{https://arxiv.org/abs/1709.09660}{{\ttfamily
  1709.09660}}].

\bibitem{LIGO2017PRL119.161101}
{LIGO Scientific Collaboration and Virgo Collaboration}, \emph{{GW170817:
  Observation of Gravitational Waves from a Binary Neutron Star Inspiral}},
  \href{https://doi.org/10.1103/PhysRevLett.119.161101}{\emph{\prl} {\bfseries
  119} (2017) 161101} [\href{https://arxiv.org/abs/1710.05832}{{\ttfamily
  1710.05832}}].

\bibitem{Agazie2023ApJL951.L8}
G.~{Agazie}, A.~{Anumarlapudi} and {Archibald}, \emph{{The NANOGrav 15 yr Data
  Set: Evidence for a Gravitational-wave Background}},
  \href{https://doi.org/10.3847/2041-8213/acdac6}{\emph{\apjl} {\bfseries 951}
  (2023) L8} [\href{https://arxiv.org/abs/2306.16213}{{\ttfamily 2306.16213}}].

\bibitem{LIGO2022PRD106.102008}
{LIGO Scientific Collaboration and Virgo Collaboration}, \emph{{All-sky search
  for continuous gravitational waves from isolated neutron stars using Advanced
  LIGO and Advanced Virgo O3 data}},
  \href{https://doi.org/10.1103/PhysRevD.106.102008}{\emph{\prd} {\bfseries
  106} (2022) 102008} [\href{https://arxiv.org/abs/2201.00697}{{\ttfamily
  2201.00697}}].

\bibitem{Bodaghee2023ApJ951.37}
A.~{Bodaghee}, J.L.~{Chiu}, J.A.~{Tomsick}, V.~{Bhalerao}, E.~{Bottacini},
  M.~{Clavel} et~al., \emph{{Drop in the Hard Pulsed Fraction and a Candidate
  Cyclotron Line in IGR J16320-4751 Seen by NuSTAR}},
  \href{https://doi.org/10.3847/1538-4357/acd541}{\emph{\apj} {\bfseries 951}
  (2023) 37} [\href{https://arxiv.org/abs/2305.07068}{{\ttfamily 2305.07068}}].

\bibitem{Weinstein2016arXiv1602.04040}
G.~{Weinstein}, \emph{{Einstein's Discovery of Gravitational Waves 1916-1918}},
  \href{https://doi.org/10.48550/arXiv.1602.04040}{\emph{arXiv e-prints} (2016)
  arXiv:1602.04040} [\href{https://arxiv.org/abs/1602.04040}{{\ttfamily
  1602.04040}}].

\bibitem{Einstein1916AnP354.769}
A.~{Einstein}, \emph{{Die Grundlage der allgemeinen Relativit{\"a}tstheorie}},
  \href{https://doi.org/10.1002/andp.19163540702}{\emph{Annalen der Physik}
  {\bfseries 354} (1916) 769}.

\bibitem{Einstein1916SPAW688}
A.~{Einstein}, \emph{{N{\"a}herungsweise Integration der Feldgleichungen der
  Gravitation}}, {\emph{Sitzungsberichte der K\&ouml;niglich Preussischen
  Akademie der Wissenschaften} (1916) 688}.

\bibitem{Einstein1918SPAW154}
A.~{Einstein}, \emph{{{\"U}ber Gravitationswellen}}, {\emph{Sitzungsberichte
  der K\&ouml;niglich Preussischen Akademie der Wissenschaften} (1918) 154}.

\bibitem{Eddington1922RSPSA102.268}
A.S.~{Eddington}, \emph{{The Propagation of Gravitational Waves}},
  \href{https://doi.org/10.1098/rspa.1922.0085}{\emph{Proceedings of the Royal
  Society of London Series A} {\bfseries 102} (1922) 268}.

\bibitem{Eddington1923Book}
A.S.~{Eddington}, \emph{{The Mathematical Theory of Relativity}}, Cambridge
  University Press, {Cambridge, England} (1923).

\bibitem{Hughes2001CQG18.4067}
S.A.~{Hughes}, \emph{{Gravitational waves from extreme mass ratio inspirals:
  challenges in mapping the spacetime of massive, compact objects}},
  \href{https://doi.org/10.1088/0264-9381/18/19/314}{\emph{Classical and
  Quantum Gravity} {\bfseries 18} (2001) 4067}
  [\href{https://arxiv.org/abs/gr-qc/0008058}{{\ttfamily gr-qc/0008058}}].

\bibitem{Tinto2005LRR8.4}
M.~{Tinto} and S.V.~{Dhurandhar}, \emph{{Time-Delay Interferometry}},
  \href{https://doi.org/10.12942/lrr-2005-4}{\emph{Living Reviews in
  Relativity} {\bfseries 8} (2005) 4}.

\bibitem{Folkner1998AIPC456}
W.M.~{Folkner}, ed., \emph{{Laser Interferometer Space Antenna : Second
  International LISA Symposium on the Detection and Observation of
  Gravitational Waves in Space : Pasadena, California, July 1998}}, vol.~456 of
  \emph{American Institute of Physics Conference Series}, AIP, Dec., 1998.

\bibitem{Babak2017PRD95.103012}
S.~{Babak}, J.~{Gair}, A.~{Sesana}, E.~{Barausse}, C.F.~{Sopuerta},
  C.P.L.~{Berry} et~al., \emph{{Science with the space-based interferometer
  LISA. V. Extreme mass-ratio inspirals}},
  \href{https://doi.org/10.1103/PhysRevD.95.103012}{\emph{\prd} {\bfseries 95}
  (2017) 103012} [\href{https://arxiv.org/abs/1703.09722}{{\ttfamily
  1703.09722}}].

\bibitem{Fan2020PRD102.063016}
H.-M.~{Fan}, Y.-M.~{Hu}, E.~{Barausse}, A.~{Sesana}, J.-d.~{Zhang}, X.~{Zhang}
  et~al., \emph{{Science with the TianQin observatory: Preliminary result on
  extreme-mass-ratio inspirals}},
  \href{https://doi.org/10.1103/PhysRevD.102.063016}{\emph{\prd} {\bfseries
  102} (2020) 063016} [\href{https://arxiv.org/abs/2005.08212}{{\ttfamily
  2005.08212}}].

\bibitem{Amaro-Seoane2017arXiv170200786}
{LISA Collaboration}, \emph{{Laser Interferometer Space Antenna}},
  \href{https://doi.org/10.48550/arXiv.1702.00786}{\emph{arXiv e-prints} (2017)
  arXiv:1702.00786} [\href{https://arxiv.org/abs/1702.00786}{{\ttfamily
  1702.00786}}].

\bibitem{Maselli2022NA6.464}
A.~{Maselli}, N.~{Franchini}, L.~{Gualtieri}, T.P.~{Sotiriou}, S.~{Barsanti}
  and P.~{Pani}, \emph{{Detecting fundamental fields with LISA observations of
  gravitational waves from extreme mass-ratio inspirals}},
  \href{https://doi.org/10.1038/s41550-021-01589-5}{\emph{\na} {\bfseries 6}
  (2022) 464} [\href{https://arxiv.org/abs/2106.11325}{{\ttfamily
  2106.11325}}].

\bibitem{Robson2019CQG36.105011}
T.~{Robson}, N.J.~{Cornish} and C.~{Liu}, \emph{{The construction and use of
  LISA sensitivity curves}},
  \href{https://doi.org/10.1088/1361-6382/ab1101}{\emph{Classical and Quantum
  Gravity} {\bfseries 36} (2019) 105011}
  [\href{https://arxiv.org/abs/1803.01944}{{\ttfamily 1803.01944}}].

\bibitem{Amaro-Seoane2012CQG29.124016}
P.~{Amaro-Seoane}, S.~{Aoudia}, S.~{Babak}, P.~{Bin{\'e}truy}, E.~{Berti},
  A.~{Boh{\'e}} et~al., \emph{{Low-frequency gravitational-wave science with
  eLISA/NGO}},
  \href{https://doi.org/10.1088/0264-9381/29/12/124016}{\emph{Classical and
  Quantum Gravity} {\bfseries 29} (2012) 124016}
  [\href{https://arxiv.org/abs/1202.0839}{{\ttfamily 1202.0839}}].

\bibitem{Luo2016CQG33.035010}
J.~{Luo}, L.-S.~{Chen}, H.-Z.~{Duan}, Y.-G.~{Gong}, S.~{Hu}, J.~{Ji} et~al.,
  \emph{{TianQin: a space-borne gravitational wave detector}},
  \href{https://doi.org/10.1088/0264-9381/33/3/035010}{\emph{\cqg} {\bfseries
  33} (2016) 035010} [\href{https://arxiv.org/abs/1512.02076}{{\ttfamily
  1512.02076}}].

\bibitem{Liu2020PRD101.103027}
S.~{Liu}, Y.-M.~{Hu}, J.-d.~{Zhang} and J.~{Mei}, \emph{{Science with the
  TianQin observatory: Preliminary results on stellar-mass binary black
  holes}}, \href{https://doi.org/10.1103/PhysRevD.101.103027}{\emph{\prd}
  {\bfseries 101} (2020) 103027}
  [\href{https://arxiv.org/abs/2004.14242}{{\ttfamily 2004.14242}}].

\bibitem{Cutler2006PRD73.042001}
C.~{Cutler} and J.~{Harms}, \emph{{Big Bang Observer and the
  neutron-star-binary subtraction problem}},
  \href{https://doi.org/10.1103/PhysRevD.73.042001}{\emph{\prd} {\bfseries 73}
  (2006) 042001} [\href{https://arxiv.org/abs/gr-qc/0511092}{{\ttfamily
  gr-qc/0511092}}].

\bibitem{DECIGO2021PTEP2021.05A105}
{DECIGO collaboration}, \emph{{Current status of space gravitational wave
  antenna DECIGO and B-DECIGO}},
  \href{https://doi.org/10.1093/ptep/ptab019}{\emph{Progress of Theoretical and
  Experimental Physics} {\bfseries 2021} (2021) 05A105}
  [\href{https://arxiv.org/abs/2006.13545}{{\ttfamily 2006.13545}}].

\bibitem{Ruan2020IJMPA35.2050075}
W.-H.~{Ruan}, Z.-K.~{Guo}, R.-G.~{Cai} and Y.-Z.~{Zhang}, \emph{{Taiji program:
  Gravitational-wave sources}},
  \href{https://doi.org/10.1142/S0217751X2050075X}{\emph{International Journal
  of Modern Physics A} {\bfseries 35} (2020) 2050075}.

\bibitem{Glampedakis2005CQG22.S605}
K.~{Glampedakis}, \emph{{Extreme mass ratio inspirals: LISA's unique probe of
  black hole gravity}},
  \href{https://doi.org/10.1088/0264-9381/22/15/004}{\emph{Classical and
  Quantum Gravity} {\bfseries 22} (2005) S605}
  [\href{https://arxiv.org/abs/gr-qc/0509024}{{\ttfamily gr-qc/0509024}}].

\bibitem{Barack2004PRD69.082005}
L.~{Barack} and C.~{Cutler}, \emph{{LISA capture sources: Approximate
  waveforms, signal-to-noise ratios, and parameter estimation accuracy}},
  \href{https://doi.org/10.1103/PhysRevD.69.082005}{\emph{\prd} {\bfseries 69}
  (2004) 082005} [\href{https://arxiv.org/abs/gr-qc/0310125}{{\ttfamily
  gr-qc/0310125}}].

\bibitem{Babak2007PRD75.024005}
S.~{Babak}, H.~{Fang}, J.R.~{Gair}, K.~{Glampedakis} and S.A.~{Hughes},
  \emph{{``Kludge'' gravitational waveforms for a test-body orbiting a Kerr
  black hole}}, \href{https://doi.org/10.1103/PhysRevD.75.024005}{\emph{\prd}
  {\bfseries 75} (2007) 024005}
  [\href{https://arxiv.org/abs/gr-qc/0607007}{{\ttfamily gr-qc/0607007}}].

\bibitem{Chua2015CQG32.232002}
A.J.K.~{Chua} and J.R.~{Gair}, \emph{{Improved analytic extreme-mass-ratio
  inspiral model for scoping out eLISA data analysis}},
  \href{https://doi.org/10.1088/0264-9381/32/23/232002}{\emph{Classical and
  Quantum Gravity} {\bfseries 32} (2015) 232002}
  [\href{https://arxiv.org/abs/1510.06245}{{\ttfamily 1510.06245}}].

\bibitem{Chua2017PRD96.044005}
A.J.K.~{Chua}, C.J.~{Moore} and J.R.~{Gair}, \emph{{Augmented kludge waveforms
  for detecting extreme-mass-ratio inspirals}},
  \href{https://doi.org/10.1103/PhysRevD.96.044005}{\emph{\prd} {\bfseries 96}
  (2017) 044005} [\href{https://arxiv.org/abs/1705.04259}{{\ttfamily
  1705.04259}}].

\bibitem{Rahman2023PRD107.024006}
M.~{Rahman} and A.~{Bhattacharyya}, \emph{{Prospects for determining the nature
  of the secondaries of extreme mass-ratio inspirals using the spin-induced
  quadrupole deformation}},
  \href{https://doi.org/10.1103/PhysRevD.107.024006}{\emph{\prd} {\bfseries
  107} (2023) 024006} [\href{https://arxiv.org/abs/2112.13869}{{\ttfamily
  2112.13869}}].

\bibitem{Rahman2024JCAP01.035}
M.~{Rahman}, S.~{Kumar} and A.~{Bhattacharyya}, \emph{{Probing astrophysical
  environment with eccentric extreme mass-ratio inspirals}},
  \href{https://doi.org/10.1088/1475-7516/2024/01/035}{\emph{\jcap} {\bfseries
  2024} (2024) 035} [\href{https://arxiv.org/abs/2306.14971}{{\ttfamily
  2306.14971}}].

\bibitem{Speri2023arXiv2307.12585}
L.~{Speri}, M.L.~{Katz}, A.J.K.~{Chua}, S.A.~{Hughes}, N.~{Warburton},
  J.E.~{Thompson} et~al., \emph{{Fast and Fourier: Extreme Mass Ratio Inspiral
  Waveforms in the Frequency Domain}},
  \href{https://doi.org/10.48550/arXiv.2307.12585}{\emph{arXiv e-prints} (2023)
  arXiv:2307.12585} [\href{https://arxiv.org/abs/2307.12585}{{\ttfamily
  2307.12585}}].

\bibitem{Zi2023PRD107.023005}
T.~{Zi}, Z.~{Zhou}, H.-T.~{Wang}, P.-C.~{Li}, J.-d.~{Zhang} and B.~{Chen},
  \emph{{Analytic kludge waveforms for extreme-mass-ratio inspirals of a
  charged object around a Kerr-Newman black hole}},
  \href{https://doi.org/10.1103/PhysRevD.107.023005}{\emph{\prd} {\bfseries
  107} (2023) 023005} [\href{https://arxiv.org/abs/2205.00425}{{\ttfamily
  2205.00425}}].

\bibitem{Zi2023PRD108.024018}
T.~{Zi} and P.-C.~{Li}, \emph{{Probing the tidal deformability of the central
  object with analytic kludge waveforms of an extreme-mass-ratio inspiral}},
  \href{https://doi.org/10.1103/PhysRevD.108.024018}{\emph{\prd} {\bfseries
  108} (2023) 024018} [\href{https://arxiv.org/abs/2303.16610}{{\ttfamily
  2303.16610}}].

\bibitem{Katz2021PRD104.064047}
M.L.~{Katz}, A.J.K.~{Chua}, L.~{Speri}, N.~{Warburton} and S.A.~{Hughes},
  \emph{{Fast extreme-mass-ratio-inspiral waveforms: New tools for millihertz
  gravitational-wave data analysis}},
  \href{https://doi.org/10.1103/PhysRevD.104.064047}{\emph{\prd} {\bfseries
  104} (2021) 064047} [\href{https://arxiv.org/abs/2104.04582}{{\ttfamily
  2104.04582}}].

\bibitem{Yang2025JCAP01.091}
S.~{Yang}, Y.-P.~{Zhang}, T.~{Zhu}, L.~{Zhao} and Y.-X.~{Liu},
  \emph{{Gravitational waveforms from periodic orbits around a
  quantum-corrected black hole}},
  \href{https://doi.org/10.1088/1475-7516/2025/01/091}{\emph{\jcap} {\bfseries
  2025} (2025) 091} [\href{https://arxiv.org/abs/2407.00283}{{\ttfamily
  2407.00283}}].

\bibitem{Sundararajan2007PRD76.104005}
P.A.~{Sundararajan}, G.~{Khanna} and S.A.~{Hughes}, \emph{{Towards adiabatic
  waveforms for inspiral into Kerr black holes: A new model of the source for
  the time domain perturbation equation}},
  \href{https://doi.org/10.1103/PhysRevD.76.104005}{\emph{\prd} {\bfseries 76}
  (2007) 104005} [\href{https://arxiv.org/abs/gr-qc/0703028}{{\ttfamily
  gr-qc/0703028}}].

\bibitem{Sundararajan2008PRD78.024022}
P.A.~{Sundararajan}, G.~{Khanna}, S.A.~{Hughes} and S.~{Drasco}, \emph{{Towards
  adiabatic waveforms for inspiral into Kerr black holes. II. Dynamical sources
  and generic orbits}},
  \href{https://doi.org/10.1103/PhysRevD.78.024022}{\emph{\prd} {\bfseries 78}
  (2008) 024022} [\href{https://arxiv.org/abs/0803.0317}{{\ttfamily
  0803.0317}}].

\bibitem{Hughes2000PRD61.084004}
S.A.~{Hughes}, \emph{{Evolution of circular, nonequatorial orbits of Kerr black
  holes due to gravitational-wave emission}},
  \href{https://doi.org/10.1103/PhysRevD.61.084004}{\emph{\prd} {\bfseries 61}
  (2000) 084004} [\href{https://arxiv.org/abs/gr-qc/9910091}{{\ttfamily
  gr-qc/9910091}}].

\bibitem{Drasco2006PRD73.024027}
S.~{Drasco} and S.A.~{Hughes}, \emph{{Gravitational wave snapshots of generic
  extreme mass ratio inspirals}},
  \href{https://doi.org/10.1103/PhysRevD.73.024027}{\emph{\prd} {\bfseries 73}
  (2006) 024027} [\href{https://arxiv.org/abs/gr-qc/0509101}{{\ttfamily
  gr-qc/0509101}}].

\bibitem{Poisson1995PRD51.5753}
E.~{Poisson} and M.~{Sasaki}, \emph{{Gravitational radiation from a particle in
  circular orbit around a black hole. V. Black-hole absorption and tail
  corrections}}, \href{https://doi.org/10.1103/PhysRevD.51.5753}{\emph{\prd}
  {\bfseries 51} (1995) 5753}
  [\href{https://arxiv.org/abs/gr-qc/9412027}{{\ttfamily gr-qc/9412027}}].

\bibitem{Amaro-Seoane2007CQG24.R113}
P.~{Amaro-Seoane}, J.R.~{Gair}, M.~{Freitag}, M.C.~{Miller}, I.~{Mandel},
  C.J.~{Cutler} et~al., \emph{{TOPICAL REVIEW: Intermediate and extreme
  mass-ratio inspirals{\textemdash}astrophysics, science applications and
  detection using LISA}},
  \href{https://doi.org/10.1088/0264-9381/24/17/R01}{\emph{Classical and
  Quantum Gravity} {\bfseries 24} (2007) R113}
  [\href{https://arxiv.org/abs/astro-ph/0703495}{{\ttfamily
  astro-ph/0703495}}].

\bibitem{Gair2007CQG24.1145}
J.~{Gair} and G.~{Jones}, \emph{{Detecting extreme mass ratio inspiral events
  in LISA data using the hierarchical algorithm for clusters and ridges
  (HACR)}}, \href{https://doi.org/10.1088/0264-9381/24/5/007}{\emph{Classical
  and Quantum Gravity} {\bfseries 24} (2007) 1145}
  [\href{https://arxiv.org/abs/gr-qc/0610046}{{\ttfamily gr-qc/0610046}}].

\bibitem{Gair2008CQG25.184031}
J.R.~{Gair}, I.~{Mandel} and L.~{Wen}, \emph{{Improved time frequency analysis
  of extreme-mass-ratio inspiral signals in mock LISA data}},
  \href{https://doi.org/10.1088/0264-9381/25/18/184031}{\emph{Classical and
  Quantum Gravity} {\bfseries 25} (2008) 184031}
  [\href{https://arxiv.org/abs/0804.1084}{{\ttfamily 0804.1084}}].

\bibitem{Peters1964PR136.1224}
P.C.~{Peters}, \emph{{Gravitational Radiation and the Motion of Two Point
  Masses}}, \href{https://doi.org/10.1103/PhysRev.136.B1224}{\emph{Physical
  Review} {\bfseries 136} (1964) 1224}.

\bibitem{Junker1992MNRAS254.146}
W.~{Junker} and G.~{Schaefer}, \emph{{Binary systems - Higher order
  gravitational radiation damping and wave emission}},
  \href{https://doi.org/10.1093/mnras/254.1.146}{\emph{\mnras} {\bfseries 254}
  (1992) 146}.

\bibitem{Forseth2016PRD93.064058}
E.~{Forseth}, C.R.~{Evans} and S.~{Hopper}, \emph{{Eccentric-orbit
  extreme-mass-ratio inspiral gravitational wave energy fluxes to 7PN order}},
  \href{https://doi.org/10.1103/PhysRevD.93.064058}{\emph{\prd} {\bfseries 93}
  (2016) 064058} [\href{https://arxiv.org/abs/1512.03051}{{\ttfamily
  1512.03051}}].

\bibitem{Gopakumar1997PRD56.7708}
A.~{Gopakumar} and B.R.~{Iyer}, \emph{{Gravitational waves from inspiraling
  compact binaries: Angular momentum flux, evolution of the orbital elements,
  and the waveform to the second post-Newtonian order}},
  \href{https://doi.org/10.1103/PhysRevD.56.7708}{\emph{\prd} {\bfseries 56}
  (1997) 7708} [\href{https://arxiv.org/abs/gr-qc/9710075}{{\ttfamily
  gr-qc/9710075}}].

\bibitem{Gopakumar1998PhRvD57.6562}
A.~{Gopakumar}, B.R.~{Iyer} and S.~{Iyer}, \emph{{Erratum: Second
  post-Newtonian gravitational radiation reaction for two-body systems:
  Nonspinning bodies [Phys. Rev. D 55, 6030 (1997)]}},
  \href{https://doi.org/10.1103/PhysRevD.57.6562}{\emph{\prd} {\bfseries 57}
  (1998) 6562}.

\bibitem{Ashtekar2002PRL89.261101}
A.~{Ashtekar} and B.~{Krishnan}, \emph{{Dynamical Horizons: Energy, Angular
  Momentum, Fluxes, and Balance Laws}},
  \href{https://doi.org/10.1103/PhysRevLett.89.261101}{\emph{\prl} {\bfseries
  89} (2002) 261101} [\href{https://arxiv.org/abs/gr-qc/0207080}{{\ttfamily
  gr-qc/0207080}}].

\bibitem{Hasselmann1963JFM15.385}
K.~{Hasselmann}, \emph{{On the non-linear energy transfer in a gravity-wave
  spectrum. Part 3. Evaluation of the energy flux and swell-sea interaction for
  a Neumann spectrum}},
  \href{https://doi.org/10.1017/S002211206300032X}{\emph{Journal of Fluid
  Mechanics} {\bfseries 15} (1963) 385}.

\bibitem{Maggiore2007gwte.book}
M.~{Maggiore}, \emph{{Gravitational Waves: Volume 1: Theory and Experiments}}
  (2007),
  \href{https://doi.org/10.1093/acprof:oso/9780198570745.001.0001}{10.1093/acprof:oso/9780198570745.001.0001}.

\bibitem{Poisson2008CQG25.209002}
E.~{Poisson}, \emph{{BOOK REVIEW: Gravitational Waves, Volume 1: Theory and
  Experiments}},
  \href{https://doi.org/10.1088/0264-9381/25/20/209002}{\emph{Classical and
  Quantum Gravity} {\bfseries 25} (2008) 209002}.

\bibitem{Thorne1980RMP52.299}
K.S.~{Thorne}, \emph{{Multipole expansions of gravitational radiation}},
  \href{https://doi.org/10.1103/RevModPhys.52.299}{\emph{\rmp} {\bfseries 52}
  (1980) 299}.

\bibitem{Blanchet2002PRD65.064005}
L.~{Blanchet}, B.R.~{Iyer} and B.~{Joguet}, \emph{{Gravitational waves from
  inspiraling compact binaries: Energy flux to third post-Newtonian order}},
  \href{https://doi.org/10.1103/PhysRevD.65.064005}{\emph{\prd} {\bfseries 65}
  (2002) 064005} [\href{https://arxiv.org/abs/gr-qc/0105098}{{\ttfamily
  gr-qc/0105098}}].

\bibitem{Blanchet2005PRD71.129903}
L.~{Blanchet}, B.R.~{Iyer} and B.~{Joguet}, \emph{{Erratum: Gravitational waves
  from inspiraling compact binaries: Energy flux to third post-Newtonian order
  [Phys. Rev. D 65, 064005 (2002)]}},
  \href{https://doi.org/10.1103/PhysRevD.71.129903}{\emph{\prd} {\bfseries 71}
  (2005) 129903}.

\bibitem{Boyle2008PRD78.104020}
M.~{Boyle}, A.~{Buonanno}, L.E.~{Kidder}, A.H.~{Mrou{\'e}}, Y.~{Pan},
  H.P.~{Pfeiffer} et~al., \emph{{High-accuracy numerical simulation of
  black-hole binaries: Computation of the gravitational-wave energy flux and
  comparisons with post-Newtonian approximants}},
  \href{https://doi.org/10.1103/PhysRevD.78.104020}{\emph{\prd} {\bfseries 78}
  (2008) 104020} [\href{https://arxiv.org/abs/0804.4184}{{\ttfamily
  0804.4184}}].

\bibitem{Thrane2013PRD88.l24032}
E.~{Thrane} and J.D.~{Romano}, \emph{{Sensitivity curves for searches for
  gravitational-wave backgrounds}},
  \href{https://doi.org/10.1103/PhysRevD.88.124032}{\emph{\prd} {\bfseries 88}
  (2013) 124032} [\href{https://arxiv.org/abs/1310.5300}{{\ttfamily
  1310.5300}}].

\bibitem{Torres2011heep}
D.F.~{Torres} and N.~{Rea}, eds., \emph{{High-Energy Emission from Pulsars and
  their Systems: Proceedings of the First Session of the Sant Cugat Forum on
  Astrophysics}}, Astrophysics and Space Science Proceedings, Springer, Berlin,
  Heidelberg (2011),
  \href{https://doi.org/10.1007/978-3-642-17251-9}{10.1007/978-3-642-17251-9}.

\bibitem{Moore2015CQG32.015014}
C.J.~{Moore}, R.H.~{Cole} and C.P.L.~{Berry}, \emph{{Gravitational-wave
  sensitivity curves}},
  \href{https://doi.org/10.1088/0264-9381/32/1/015014}{\emph{Classical and
  Quantum Gravity} {\bfseries 32} (2015) 015014}
  [\href{https://arxiv.org/abs/1408.0740}{{\ttfamily 1408.0740}}].

\bibitem{Antoniadis2023A&A678.A50}
{EPTA Collaboration}, {InPTA Collaboration}, J.~{Antoniadis}, P.~{Arumugam},
  S.~{Arumugam}, S.~{Babak} et~al., \emph{{The second data release from the
  European Pulsar Timing Array. III. Search for gravitational wave signals}},
  \href{https://doi.org/10.1051/0004-6361/202346844}{\emph{\aap} {\bfseries
  678} (2023) A50} [\href{https://arxiv.org/abs/2306.16214}{{\ttfamily
  2306.16214}}].

\bibitem{Hobbs2010CQG27.084013}
G.~{Hobbs}, A.~{Archibald}, Z.~{Arzoumanian}, D.~{Backer}, M.~{Bailes},
  N.D.R.~{Bhat} et~al., \emph{{The International Pulsar Timing Array project:
  using pulsars as a gravitational wave detector}},
  \href{https://doi.org/10.1088/0264-9381/27/8/084013}{\emph{Classical and
  Quantum Gravity} {\bfseries 27} (2010) 084013}
  [\href{https://arxiv.org/abs/0911.5206}{{\ttfamily 0911.5206}}].

\bibitem{Braun2015SKA174}
R.~{Braun}, T.~{Bourke}, J.A.~{Green}, E.~{Keane} and J.~{Wagg},
  \emph{{Advancing Astrophysics with the Square Kilometre Array}},  in
  \emph{Advancing Astrophysics with the Square Kilometre Array (AASKA14)},
  p.~174, Apr., 2015, \href{https://doi.org/10.22323/1.215.0174}{DOI}.

\bibitem{LIGO2017CQG34.044001}
{LIGO Scientific Collaboration}, \emph{{Exploring the sensitivity of next
  generation gravitational wave detectors}},
  \href{https://doi.org/10.1088/1361-6382/aa51f4}{\emph{Classical and Quantum
  Gravity} {\bfseries 34} (2017) 044001}
  [\href{https://arxiv.org/abs/1607.08697}{{\ttfamily 1607.08697}}].

\bibitem{LIGO2021PRX11.021053}
{LIGO Scientific Collaboration and Virgo Collaboration}, \emph{{GWTC-2: Compact
  Binary Coalescences Observed by LIGO and Virgo during the First Half of the
  Third Observing Run}},
  \href{https://doi.org/10.1103/PhysRevX.11.021053}{\emph{Physical Review X}
  {\bfseries 11} (2021) 021053}
  [\href{https://arxiv.org/abs/2010.14527}{{\ttfamily 2010.14527}}].

\bibitem{Tso2022Aplus}
R.~Tso and the LIGO A+~Collaboration, \emph{{The A+ upgrade: Improving
  low-frequency sensitivity in Advanced LIGO}},
  \href{https://doi.org/10.1088/1361-6382/ac728a}{\emph{Classical and Quantum
  Gravity} {\bfseries 39} (2022) 155001}.

\bibitem{Bonanno2000PRD62.043008}
A.~{Bonanno} and M.~{Reuter}, \emph{{Renormalization group improved black hole
  spacetimes}}, \href{https://doi.org/10.1103/PhysRevD.62.043008}{\emph{\prd}
  {\bfseries 62} (2000) 043008}
  [\href{https://arxiv.org/abs/hep-th/0002196}{{\ttfamily hep-th/0002196}}].

\bibitem{Scardigli2023PRD107.104001}
F.~{Scardigli} and G.~{Lambiase}, \emph{{Planck stars from a scale-dependent
  gravity theory}},
  \href{https://doi.org/10.1103/PhysRevD.107.104001}{\emph{\prd} {\bfseries
  107} (2023) 104001} [\href{https://arxiv.org/abs/2205.07088}{{\ttfamily
  2205.07088}}].

\bibitem{Bonanno2002PRD65.043508}
A.~{Bonanno} and M.~{Reuter}, \emph{{Cosmology of the Planck era from a
  renormalization group for quantum gravity}},
  \href{https://doi.org/10.1103/PhysRevD.65.043508}{\emph{\prd} {\bfseries 65}
  (2002) 043508} [\href{https://arxiv.org/abs/hep-th/0106133}{{\ttfamily
  hep-th/0106133}}].

\bibitem{Huang2024PRD109.l24005}
L.~{Huang} and X.-M.~{Deng}, \emph{{Can a particle's motion distinguish
  scale-dependent Planck stars from renormalization group improved
  Schwarzschild black holes?}},
  \href{https://doi.org/10.1103/PhysRevD.109.124005}{\emph{\prd} {\bfseries
  109} (2024) 124005}.

\bibitem{Lambiase2022PhRvD105.124054}
G.~{Lambiase} and F.~{Scardigli}, \emph{{Generalized uncertainty principle and
  asymptotically safe gravity}},
  \href{https://doi.org/10.1103/PhysRevD.105.124054}{\emph{Physical Review D}
  {\bfseries 105} (2022) 124054}
  [\href{https://arxiv.org/abs/2204.07416}{{\ttfamily 2204.07416}}].

\bibitem{Fujita2009CQG26.135002}
R.~{Fujita} and W.~{Hikida}, \emph{{Analytical solutions of bound timelike
  geodesic orbits in Kerr spacetime}},
  \href{https://doi.org/10.1088/0264-9381/26/13/135002}{\emph{Classical and
  Quantum Gravity} {\bfseries 26} (2009) 135002}
  [\href{https://arxiv.org/abs/0906.1420}{{\ttfamily 0906.1420}}].

\bibitem{Newton1687book}
I.~{Newton}, \emph{{Philosophiae Naturalis Principia Mathematica.}} (1687),
  \href{https://doi.org/10.3931/e-rara-440}{10.3931/e-rara-440}.

\bibitem{Brouwer1961Book}
D.~{Brouwer} and G.M.~{Clemence}, \emph{{Methods of Celestial Mechanics}},
  Academic Press, New York (1961).

\bibitem{Laskar1989Nat338.237}
J.~{Laskar}, \emph{{A numerical experiment on the chaotic behaviour of the
  Solar System}}, \href{https://doi.org/10.1038/338237a0}{\emph{\nat}
  {\bfseries 338} (1989) 237}.

\bibitem{Mayor1995Nat378.355}
M.~{Mayor} and D.~{Queloz}, \emph{{A Jupiter-mass companion to a solar-type
  star}}, \href{https://doi.org/10.1038/378355a0}{\emph{\nat} {\bfseries 378}
  (1995) 355}.

\bibitem{Liu2024JCAP10.056}
Y.~{Liu} and X.~{Zhang}, \emph{{Gravitational waves for eccentric extreme mass
  ratio inspirals of self-dual spacetime}},
  \href{https://doi.org/10.1088/1475-7516/2024/10/056}{\emph{\jcap} {\bfseries
  2024} (2024) 056} [\href{https://arxiv.org/abs/2404.08454}{{\ttfamily
  2404.08454}}].

\bibitem{Huang2025PRD111.084038}
L.~{Huang}, \emph{{Probing holonomy corrected Schwarzschild black holes with
  precessing and periodic orbits}},
  \href{https://doi.org/10.1103/PhysRevD.111.084038}{\emph{\prd} {\bfseries
  111} (2025) 084038}.

\bibitem{Deng2021PDU31.100745}
H.-Y.~{Lin} and X.-M.~{Deng}, \emph{{Rational orbits around 4 D
  Einstein-Lovelock black holes}},
  \href{https://doi.org/10.1016/j.dark.2020.100745}{\emph{\pdu} {\bfseries 31}
  (2021) 100745}.

\bibitem{Deng2023EPJC83.311}
H.-Y.~{Lin} and X.-M.~{Deng}, \emph{{Precessing and periodic orbits around
  hairy black holes in Horndeski's Theory}},
  \href{https://doi.org/10.1140/epjc/s10052-023-11487-x}{\emph{\epjc}
  {\bfseries 83} (2023) 311}.

\bibitem{Levin2008PRD77.103005}
J.~{Levin} and G.~{Perez-Giz}, \emph{{A periodic table for black hole orbits}},
  \href{https://doi.org/10.1103/PhysRevD.77.103005}{\emph{\prd} {\bfseries 77}
  (2008) 103005} [\href{https://arxiv.org/abs/0802.0459}{{\ttfamily
  0802.0459}}].

\bibitem{Levin2009PRD79.043016}
J.~{Levin} and R.~{Grossman}, \emph{{Dynamics of black hole pairs. I. Periodic
  tables}}, \href{https://doi.org/10.1103/PhysRevD.79.043016}{\emph{\prd}
  {\bfseries 79} (2009) 043016}
  [\href{https://arxiv.org/abs/0809.3838}{{\ttfamily 0809.3838}}].

\bibitem{Poisson2014Book}
E.~{Poisson} and C.M.~{Will}, \emph{{Gravity: Newtonian, Post-Newtonian,
  Relativistic}}, {Cambridge, England}, {Cambridge University Press} (May,
  2014),
  \href{https://doi.org/10.1017/CBO9781139507486}{10.1017/CBO9781139507486}.

\bibitem{Finn2000PRD62.l24021}
L.S.~{Finn} and K.S.~{Thorne}, \emph{{Gravitational waves from a compact star
  in a circular, inspiral orbit, in the equatorial plane of a massive, spinning
  black hole, as observed by LISA}},
  \href{https://doi.org/10.1103/PhysRevD.62.124021}{\emph{\prd} {\bfseries 62}
  (2000) 124021} [\href{https://arxiv.org/abs/gr-qc/0007074}{{\ttfamily
  gr-qc/0007074}}].

\end{thebibliography}\endgroup

\end{document}